\documentclass[aps,prd,twocolumn,floats,floatfix,showpacs,superscriptaddress,nofootinbib]{revtex4-2}

\usepackage{graphicx,amssymb,amsmath,amsthm,amsfonts,epsfig,epsf,fixmath}
\usepackage[usenames]{color}
\usepackage{epstopdf}
\usepackage{booktabs}
\usepackage{aas_macros}
\usepackage{bm}
\usepackage{dcolumn}
\usepackage[dvipsnames]{xcolor}
\usepackage[colorlinks]{hyperref}
\usepackage{cleveref}
\usepackage[utf8]{inputenc}
\usepackage{resizegather}

\begin{document}

\title{Quasi-normal modes of rotating black holes in Einstein-dilaton Gauss-Bonnet gravity: the second order in rotation}

\author{Lorenzo Pierini}
	\email{lorenzo.pierini@uniroma1.it}
	\affiliation{Dipartimento di Fisica, ``Sapienza'' Universit\`a di Roma \& Sezione INFN Roma1, P. Aldo Moro 
5, 00185 Roma, Italy}
\author{Leonardo Gualtieri}
	\email{leonardo.gualtieri@unipi.it}
	\affiliation{Dipartimento di Fisica, Universit\`a di Pisa \& Sezione INFN Pisa, L. Bruno Pontecorvo
3, 56127 Pisa, Italy}

        \begin{abstract}
  One of the most promising strategies to test gravity in the strong-field, large curvature regime is gravitational
  spectroscopy: the measurement of black hole quasi-normal modes from the ringdown signal emitted in the aftermath of a
  compact binary coalescence, searching for deviations from the predictions of general relativity.  This strategy is
  only effective if we know how quasi-normal modes of black holes are affected by modifications of general relativity;
  and if we know this for rotating black holes, since binary coalescences typically lead to black holes with spins
  $J/M^2\sim 0.7$. In this article, we compute for the first time the gravitational quasi-normal modes of rotating black
  holes up to second order in the spin in a modified gravity theory. We consider Einstein-dilaton Gauss-Bonnet gravity,
  one of the simplest theories which modifies the large-curvature regime of gravity and which can be tested with black
  hole observations. To enhance the domain of validity of the spin expansion, we perform a Pad
  \'e\'e
  resummation of the
  quasi-normal modes. We find that when the second order in spin is not included, the effect of gravity modifications
  may be seriously underestimated.  A comparison with the general relativistic case suggests that this approach should
  be accurate up to spins $\sim 0.7$; therefore, our results can be used in the data analysis of ringdown signals.
\end{abstract}

\maketitle

\section{Introduction} \label{sec:intro}
In the last century, a plethora of observations confirmed that the gravitational interaction is well described by
general relativity (GR)\,\cite{Will:2014kxa}. However, before 2015 -- when the first gravitational wave (GW) signal was
detected\,\cite{Abbott:2016blz} -- these tests were mostly limited to the weak-field regime of gravity. In recent years,
the observations of the Advanced LIGO/Virgo detectors started to explore strong gravitational fields and large spacetime
curvature, and have excluded large deviations of GR in this regime\,\cite{LIGOScientific:2021sio}. The next generation of
GW detectors, like the ground-based Einstein Telescope\,\cite{Punturo:2010zz}, will be sensitive enough to detect even
tiny GR deviations in the strong-field, large curvature regime of gravity.

One of the most promising strategies to find such deviations is the analysis of the ringdown signal emitted in the
aftermath of a binary black hole (BH) coalescence. In this stage, the waveform is a superposition of damped oscillations, the
quasi-normal modes (QNMs) of the final BH~\cite{Kokkotas:1999bd,Ferrari:2007dd,Berti:2009kk}. The QNM frequencies and
damping times carry the imprint of the underlying theory of gravity; once measured from the GW signal, they can be
compared with the predictions of GR, or - if a deviation is observed - with the predictions of a possible modified
theory of gravity. With a more refined analysis, it is possible to improve the accuracy by stacking multiple
observations\,\cite{Meidam:2014jpa,Yang:2017zxs,Maselli:2019mjd}.

This approach, which has been called ``gravitational spectroscopy''~\cite{Dreyer:2003bv,Berti:2005ys,Berti:2007zu} (see
also\,\cite{Berti:2018vdi} and references therein), requires the knowledge of the QNMs of BHs in modified gravity
theories. The frequencies and damping times of BH QNMs have been computed for a certain number of such theories, for
static, non-rotating
BHs\,\cite{Cardoso:2009pk,Molina:2010fb,Kobayashi:2012kh,Kobayashi:2014wsa,Salcedo2016,Tattersall:2019nmh,Blazquez-Salcedo:2020caw,Blazquez-Salcedo:2017txk}
or for slowly rotating BHs at first order in the
spin\,\cite{Pierini:2021jxd,Wagle:2021tam,Srivastava:2021imr,Cano:2021myl}.\footnote{An alternative approach to study BH
  QNMs in modified gravity consists in the modification of the radial potential in the perturbation equations, computing
  how these deformations affect the QNM frequencies and damping
  times~\cite{Cardoso:2019mqo,McManus:2019ulj,Volkel:2022aca}.}

However, since BH remnants of compact binary coalescences have typical spins -- for comparable-mass binaries -- of the
order of $\bar a=J/M^2\sim0.7$ (where $M,J$ are the mass and angular momentum of the BH, respectively), the knowledge of
BH QNMs to $O(\bar a)$ is inadequate to perform gravitational spectroscopy.
Indeed, an analysis of the QNMs in GR suggests that the slowly rotating approximation at ${\cal O}(\bar a)$ gives the
QNM frequencies with an accuracy of $1\%$ for $\bar a\lesssim0.2$ only (see Fig.\,\ref{fig:delta-kerr-SR}
below).\footnote{A similar computation in\,\cite{Wagle:2021tam} leads to a smaller agreement. This is mainly due to the
  fact that, after the computation of the mode frequencies $\omega^{nl}(\bar a)$ by solving the perturbation equations
  at first order in the spin, we perform a Taylor expansion of $\omega^{nl}(\bar a)$ around $\bar a=0$ (see
  Sec.~\ref{sec:secondorder}); we find that this improves the accuracy of the modes in the slow-rotation regime.} This
article, to our knowledge, is the first computation of the QNMs of a slowly rotating BH {\it to second order in the
  spin} in a modified gravity theory. We shall argue that, with an appropriate resummation of the spin expansion, this
computation is expected to be accurate for spins as large as $\bar a\sim0.7$.

We shall consider Einstein-dilaton Gauss-Bonnet (EdGB) gravity, one of the simplest theories which modifies the
strong-field, large-curvature regime of gravity, and which can be tested with BH observations (see\,\cite{TopicalReview}
and references therein). In EdGB gravity the gravitational sector contains, besides the metric tensor, a scalar field
(i.e., it is a {\it scalar-tensor theory}), which is coupled with the spacetime curvature through the Gauss-Bonnet term
$\mathcal{R}^2_{\rm GB}=R_{\mu\nu\rho\sigma} R^{\mu\nu\rho\sigma} -4 R_{\mu\nu} R^{\mu\nu} +R^2$, where
$R_{\mu\nu\alpha\beta}$, $R_{\mu\nu}$, $R$ are the Riemann tensor, the Ricci tensor and the Ricci scalar, respectively.
The action
is\,\cite{Mignemi:1992nt,Kanti1995}:
\begin{align}
  S &= \int d^4 x \frac{\sqrt{-g}}{16 \pi} \left(R-\frac{1}{2} \partial_\mu \phi \partial^\mu \phi +V(\phi)
  + \frac{\alpha_{\rm GB}}{4} e^\phi \mathcal{R}^2_{\rm GB} \right)\nonumber\\
  &+S_{\rm m}
\label{eq:S_edgb}
\end{align}
where $S_{\rm m}$ is the matter Lagrangian, which we do not consider in this paper since we are interested in BH
spacetimes. We also neglect, for simplicity, the scalar field potential $V(\phi)$.

EdGB gravity naturally arises in low-energy truncations of string theories; it belongs to the class of Horndeski
gravity~\cite{Horndeski:1974wa,Kobayashi:2019hrl}, i.e. the scalar-tensor theories with second-order-in-time field
equations (which are thus free from the Ostrogradsky instability). At variance with several other scalar-tensor
theories, EdGB gravity does not satisfy the no-hair theorems: stationary BHs have a non-trivial scalar field profile,
and are not described by the Kerr metric.

As discussed in Sec.~\ref{sec:stationaryebhdgb}, a static BH with mass $M$ can exist in EdGB gravity only if
$\alpha_{\rm GB}/M^2\lesssim 0.69$; a similar bound applies for stationary, rotating BHs. Therefore, the existence of
the lightest BH observed, J1655-40, with mass $M\simeq5.4\,M_\odot$ implies~$\sqrt{\alpha_{\rm GB}}<6.6$ Km. Current
observations of binary BH coalescences by LIGO and Virgo lead to a comparable constraint, $\sqrt{\alpha_{\rm GB}}<9.1$
Km\,\cite{Perkins:2021mhb}.\footnote{Note that the different conventions in this article and in\,\cite{Perkins:2021mhb} lead to a
correction factor of $4\,^4\sqrt{\pi}$ in the definition of $\sqrt{\alpha_{\rm GB}}$, see e.g.\,\cite{Witek:2018dmd}.}

QNMs of BHs in EdGB gravity have been computed in~\cite{Salcedo2016} for non-rotating BHs, and
in\,\cite{Pierini:2021jxd} (hereafter, Paper I) for rotating BHs at first order in the spin.  In this article we shall
compute the QNMs of stationary, rotating BHs in EdGB gravity, by performing a {\it slow-rotation expansion}, as
in~\cite{Hartle2,Pani2012scalar,Pani:2013ija}, to second order in the spin $\bar a$.

We shall use geometric units, $G=c=1$. In Section~\ref{sec:stationaryebhdgb} we describe stationary BHs in EdGB
gravity. In Section~\ref{sec:perturbations} we discuss perturbations of the stationary BH background, up to second order
in $\bar a$. In Section~\ref{sec:results} we discuss the results of our computations, and in Section~\ref{sec:concl} we
draw our conclusions. In the Appendix we give further details on the perturbation equations to second order in the spin.

\section{Stationary black holes in  Einstein-dilaton Gauss-Bonnet gravity} \label{sec:stationaryebhdgb}
The field equations obtained from~\eqref{eq:S_edgb} (with $V(\phi)=S_m=0$) are
\begin{align}
  \label{eq:scalar}
  &\frac{1}{\sqrt{-g}}\partial_\mu (\sqrt{-g} g^{\mu \nu} \partial_\nu \phi) =
  \frac{\alpha_{\rm GB}}{4} e^\phi \mathcal{R}^2_{GB}\\
 \label{eq:metric}&G_{\mu \nu}=\frac{1}{2} \partial_\mu \phi \partial_\nu \phi -
\frac{1}{4} g_{\mu\nu} (\partial_\rho \phi) (\partial^\rho \phi) - \alpha_{\rm GB} \mathcal{K}_{\mu\nu}
\end{align}
where $G_{\mu\nu}$ is the Einstein tensor,
\begin{equation}
  \mathcal{K}_{\mu\nu}= \frac{1}{8}(g_{\mu\rho} g_{\nu\sigma} + g_{\mu\sigma} g_{\nu\rho})
  \epsilon^{\delta \sigma \gamma \alpha} \nabla_\beta \left({\tilde{R}^{\rho \beta}}_{ ~~\gamma \alpha}e^\phi \partial_\delta\phi\right)\,,
\label{eq:tensore_K}
\end{equation}
$\epsilon^{\mu\nu\delta\gamma}$ is the Levi-Civita tensor, and ${\tilde{R}^{\mu \nu}}_{ ~~\rho \sigma} =
\epsilon^{\mu\nu \delta \gamma}R_{\delta \gamma \rho \sigma}$.
\subsection{Non-rotating black holes}
The solution of Eqs.~\eqref{eq:scalar}, \eqref{eq:metric} describing a static, spherically symmetric BH has 
been derived in~\cite{Kanti1995} (see also~\cite{PaniCardoso2009}). In this case, the spacetime metric can be written
as:
\begin{equation}
\label{eq:staticmetric}ds^2=-A(r)dt^2+\frac{dr^2}{B(r)}+r^2d\Omega^2
\end{equation}
where $d\Omega^2=d\theta^2+\sin^2\theta d\varphi^2$. The functions $A(r)$, $B(r)$ and $\phi(r)$ are found by numerical
integration\,\cite{Kanti1995,PaniCardoso2009}. By requiring asymptotic flatness and an asymptotically
vanishing scalar field, one finds an unique solution, with an ADM mass $M$ and a scalar charge $D$ which appear in the
asymptotic expansion of the metric and of the scalar field:
\begin{align}
  A&=1-\frac{2M}{r}+O\left(\frac{1}{r^3}\right)\nonumber\\
  \phi&=\frac{D}{r}+\frac{DM}{r^2}+O\left(\frac{1}{r^3}\right)\,.
\end{align}
The solution depends on $\alpha_{\rm GB}$ through the dimensionless coupling parameter
\begin{equation}
  \zeta=\frac{\alpha_{\rm GB}}{M^2}\,,
  \label{eq:zeta}
\end{equation}
which has to satisfy the condition
\begin{equation}
\label{eq:zetacond}  0\le\zeta<\zeta_{\rm max}\simeq 0.691\,.
\end{equation}
If $\zeta\ge\zeta_{\rm max}$, it is not possible to enforce regular boundary conditions on the horizon, and the BH
becomes a naked singularity\,\cite{Sotiriou:2014pfa}.

The scalar charge $D$ and the horizon radius $r_{\rm h}$ depend on the BH mass and of the coupling $\zeta$:
\begin{align}
  \frac{D}{M}&=\frac{\zeta}{2}+\frac{73}{60}\zeta^2+O(\zeta^3)\nonumber\\
  \frac{r_{\rm h}}{2M}&=1-\frac{49}{1280}\,\zeta^2+O(\zeta^3)\,.
\end{align}
For a given value of the coupling constant $\alpha_{\rm GB}$, there is a single static, spherically symmetric BH solution for each value of the mass satisfying $M\ge M_{\rm min}=\alpha_{\rm GB}\zeta^{-1/2}_{\rm max}$.

The field equations~\eqref{eq:scalar}, \eqref{eq:metric} can also be solved perturbatively in the  parameter
$\zeta$~\cite{Yunes:2011we}. This leads to an analytic expression for the metric and the scalar field:
\begin{align}
A(r)&=1-\frac{2M}{r}+\sum_{j=2}^{N_\zeta}\zeta^jA^{(j)}(r)\nonumber\\
B(r)&=1-\frac{2M}{r}+\sum_{j=2}^{N_\zeta}\zeta^jB^{(j)}(r)\nonumber\\
\phi(r)&=\sum_{j=1}^{N_\zeta}\zeta^j\phi^{(j)}(r)\label{eq:expABphi}
\end{align}
where $A^{(j)}(r)$, $B^{(j)}(r)$ and $\phi^{(j)}(r)$ can be written as expansions in powers of $1/r$.
%
\subsection{Rotating black holes}\label{subsec:eqrot}
%
In the case of stationary, rotating BHs, Eqs.~\eqref{eq:scalar}, \eqref{eq:tensore_K} have been solved numerically in
Refs.~\cite{Kleihaus:2011tg,Kleihaus:2015aje}, with no assumptions on the rotation rate. They have also been solved
analytically~\cite{PaniCardoso2009,Maselli} in terms of a perturbative expansion in the coupling parameter $\zeta$ and
in the spin ${\bar a}$:
\begin{align}\label{eq:metricslowrot}
  ds^2 &= - A(r)[1+2h(r,\theta)] dt^2 + \frac{1}{B(r)}[1+2p(r,\theta)]dr^2 \nonumber\\
  &+ r^2 [1+2k(r,\theta)] \left[d\theta^2+\sin^2\theta(d\varphi-\varpi(r,\theta) dt)^2\right]
\end{align}
where $A(r)$, $B(r)$, $\phi(r)$ are given in Eqs.~\eqref{eq:expABphi}, while $\varpi(r,\theta)$, $h(r,\theta)$, $p(r,\theta)$
and the scalar field $\phi(r,\theta)$ are given as expansions in $\zeta$, $\bar a$ and in the Legendre polynomials $P_l(\theta)$.
For instance, the scalar field expansion is:
\begin{equation}
\phi(r)=\sum_{j=1}^{N_\zeta}\,\,\sum_{n=0,2,4,...}^{N_{\bar a}} \,\,\sum_{l =0,2,4...}^{n}\zeta^j
  {\bar a}^n \phi_l ^{(nj)}(r) P_l (\theta)\label{eq:scalarslowrot}\,,
\end{equation}
where $N_\zeta$, $N_a$, $n$ are the truncation orders of the expansions in the coupling, in the spin and in the Legendre
polynomials. The expansions of $\varpi(r,\theta)$, $h(r,\theta)$, $p(r,\theta)$ have the same structure. Note that for
$\zeta=0$ (i.e., in GR) the metric~\eqref{eq:metricslowrot} reduces to the Hartle-Thorne metric~\cite{Hartle2}.

The explicit expression of this expansion is given in~\cite{Maselli} and in the {\sc Mathematica} notebook in the
Supplemental Material~\cite{notebook}.
Similarly, the horizon radius $r_{\rm h}$, the scalar charge $D$ and the maximum allowed coupling $\zeta_{\rm max}$ acquire
corrections with respect to the non-rotating case, and can be expressed as  expansions in $\zeta$ and ${\bar a}$.

In the following we shall study perturbations around the slowly rotating BH background given by the
expansion~\eqref{eq:metricslowrot}, \eqref{eq:scalarslowrot}. We shall truncate the spin expansion at $N_{\bar a}=2$. As
discussed in Sec.~\ref{subsec:truncation}, an assessment of the slow-rotation expansion in GR, and an analysis of the
truncation error at different orders in the coupling, suggest that in this way we should be able to compute the QNMs,
for
$\bar a\sim 0.4$, with truncation errors within $\sim 1\%$, and for $\bar a\sim 0.7$, with truncation errors within
$\sim 5\%$.  Moreover, we shall truncate the expansion in the coupling at $N_\zeta=6$; this should lead to errors within
$\sim 1\%$ for coupling constant $\zeta\sim0.4$ for the real part of the QNMs, and for $\zeta\sim0.3$ for the imaginary
part.
\section{Perturbations of Einstein-dilaton Gauss-Bonnet black holes}\label{sec:perturbations}
We consider a perturbed stationary, rotating BH.  The spacetime metric and the scalar field are
\begin{align}
  g_{\mu\nu}&=g^{(0)}_{\mu\nu}+h_{\mu\nu}^{\rm pol}+h_{\mu\nu}^{\rm ax}\nonumber\\
  \phi&=\phi^{(0)}+\delta\phi\label{eq:defpert}
\end{align}
where $g^{(0)}_{\mu\nu}$, $\phi^{(0)}$ are given by Eqs.~\eqref{eq:metricslowrot}, \eqref{eq:scalarslowrot}.
The metric perturbation is decomposed in components with polar and axial parities.

\subsection{General structure of the equations}\label{sec:geneq}
The perturbations of the metric tensor and of the scalar field are expanded in tensor spherical
harmonics as: 
\begin{align}
  &\delta\phi(t,r,\theta,\varphi)=\frac{1}{r}\Phi^{lm}(r)Y^{lm}(\theta,\varphi)e^{-i\omega t}\,,\label{eq:exp_scal}\\
  &h^{\rm pol}_{\mu\nu}(t,r,\theta,\varphi)dx^\mu dx^\nu=\nonumber\\
  &\left[A(r)H_0^{lm}(r)dt^2+2H_1^{lm}(r)dtdr+B^{-1}(r)H_2^{lm}(r)dr^2\right.\nonumber\\
    &\left.+K^{lm}(r)(dr^2+\sin^2\theta d\varphi^2)\right]Y^{lm}(\theta,\varphi)e^{-i\omega t}\,,\label{eq:exp_pol}\\
&h^{\rm ax}_{\mu\nu}dx^\mu dx^\nu=2(h_0^{lm}(r)dt+h_1^{lm}(r)dr)\nonumber\\
  &\times(S_{\theta}(\theta,\varphi)d\theta+S_\varphi(\theta,\varphi)d\varphi)e^{-i\omega t}\label{eq:exp_ax}
\end{align}
where we have chosen the Regge-Wheeler gauge~\cite{RW1957,Zerilli1970}, and $(S^{lm}_\theta,S^{lm}_\varphi)=
(-(\sin\theta)^{-1}Y^{lm}_{,\varphi},\sin\theta Y^{lm}_{,\theta})$.
Replacing this expansion in Eqs.~\eqref{eq:scalar}, \eqref{eq:metric} leads to a set of partial
differential equations in $r$ and $\theta$ (see App.\,\ref{app:eq}). Due to the stationarity and axial symmetry of
the background, the dependence on on $t$ and $\varphi$ factors out as $\sim e^{i(m\varphi-\omega t)}$, and the equations
with different values of $m,\omega$ are decoupled.

Following e.g.~\cite{Kojima1992} (see also~\cite{AdvancedMethods} and Paper I), we can reduce the perturbation equations
to a system of ordinary differential equations in $r$, up to second order in the spin. Since the background is not
spherically symmetric, equations with different values of $l$ are coupled. Schematically, the general structure of the
perturbation equations can be written as:
\begin{align}
0 = & {\cal{P}}_{l\,m}+ {\bar a} \,m\, \bar{\cal{P}}_{l\,m} + {\bar a}^2  \hat{\cal{P}}_{l\,m}+ \, m^2\, {\bar a}^2  {\bar{\bar{\cal{P}}}}_{l\,m} \notag \\
& +{\bar a}\left(Q_{l\,m} \tilde{\cal{A}}_{l-1\,m}+Q_{l+1\,m}\tilde{\cal{A}}_{l+1\,m}\right) \notag\\
&+ {\bar a}^2\left(Q_{l-1\,m}Q_{l\,m} \breve{\cal{P}}_{l-2\,m}+Q_{l+1\,m}Q_{l+2\,m}\breve{\cal{P}}_{l+2\,m}\right) \notag\\
&+ m \,{\bar a}^2\left(Q_{l\,m} \check{\cal{A}}_{l-1\,m}+ Q_{l+1\,m}\check{\cal{A}}_{l+1\,m}\right)\label{generalstructure-pol}\\
0 = & {\cal{A}}_{l\,m}+ {\bar a} \,m\, \bar{\cal{A}}_{l\,m} + {\bar a}^2  \hat{\cal{A}}_{lm}+ \, m^2\, {\bar a}^2  {\bar{\bar{\cal{A}}}}_{l\,m} \notag \\
& +{\bar a}\left(Q_{l\,m} \tilde{\cal{P}}_{l-1\,m}+Q_{l+1\,m}\tilde{\cal{P}}_{l+1\,m}\right) \notag\\
&+ {\bar a}^2\left(Q_{l-1\,m}Q_{l\,m} \breve{\cal{A}}_{l-2\,m}+Q_{l+1\,m}Q_{l+2\,m}\breve{\cal{A}}_{l+2\,m}\right) \notag\\
&+ m \,{\bar a}^2\left(Q_{l\,m} \check{\cal{P}}_{l-1\,m}+ Q_{l+1\,m}\check{\cal{P}}_{l+1\,m}\right)\label{generalstructure-ax}
\end{align}
where $Q_{lm}$ are constant coefficients, and ${\cal P}_{lm}$, $\bar{\cal P}_{lm}$, $\hat{\cal P}_{lm}$,
${\bar{\bar{\cal{P}}}}_{l\,m}$, $\tilde{\cal{P}}_{l\,m}$, $\breve{\cal{P}}_{l\,m}$, $\check{\cal{P}}_{l\,m}$, (${\cal
  A}_{lm}$, $\bar{\cal A}_{lm}$, $\hat{\cal A}_{lm}$, ${\bar{\bar{\cal{A}}}}_{l\,m}$, $\tilde{\cal{A}}_{l\,m}$,
$\breve{\cal{A}}_{l\,m}$, $\check{\cal{A}}_{l\,m}$) are combinations of the polar perturbation functions $H_0^{lm}$,
$H_1^{lm}$, $H_2^{lm}$, $K^{lm}$, $\Phi^{lm}$ (of the axial perturbation functions $h_0^{lm}$, $h_1^{lm}$).  We remark
that the expressions ${\cal P}_{lm}$, etc. do not depend explicitly on the harmonic index $m$.  For further details, see
Appendix~\ref{app:eq} and the Supplemental Material~\cite{notebook}.

While at zero-th order in the spin the equations (${\cal P}_{lm}=0$, ${\cal A}_{lm}=0$) are decoupled, at first order in
the spin polar perturbations with harmonic index $l$ are coupled to axial perturbations with harmonic indexes $l\pm1$,
and vice versa. Moreover, when $m\neq0$ polar (axial) perturbations are coupled to perturbations having the same $l$ and
the same parity (see the discussion in Paper I). At second order in the spin, perturbations with harmonic index $l$ are
also coupled to perturbations with same parities and harmonic indexes $l\pm2$, and (when $m\neq0$) to perturbations with
opposite parities and harmonic indexes $l\pm1$.

\subsection{Quasi-normal modes}\label{sec:qnm}

The QNMs are the proper modes at which BHs (or compact stars) oscillate when excited by non-radial perturbations (see
e.g.~\cite{Kokkotas:1999bd,Ferrari:2007dd,Berti:2009kk} and references therein). At variance with normal modes, the QNMs
are damped oscillations, since they are associated to GW emission. Therefore the corresponding frequencies $\omega$ (see
Eqs.~\eqref{eq:exp_scal}-\eqref{eq:exp_ax}) are complex: $\omega=\omega_R+i\,\omega_I$, with $\omega_I<0$ being the
inverse of the damping time of the oscillation. 

To find the QNM frequencies, we solve the perturbation equations
with Sommerfeld boundary conditions, i.e. outgoing waves at infinity and ingoing waves at the horizon.
At the horizon and at infinity, the scalar ($\Phi^{lm}(r)$) and gravitational ($H^{lm}_1(r)/r$, $K^{lm}(r)$, etc.)
perturbation functions behave as
\begin{align}
&A_{\rm in}^{lm}e^{-ik_{\rm H}r_*}+A_{\rm out}^{lm}e^{ik_{\rm H}r_*}~\,~~~(r\to r_{\rm h})\nonumber\\
&A_{\rm in}^{lm}e^{-i\omega r_*}+A_{\rm out}^{lm}e^{i\omega r_*}~~~~~~~(r\to \infty)\,,\label{eq:bc0}
\end{align}
where $r_*$ is a properly defined tortoise coordinate for the background spacetime $g^{(0)}_{\mu\nu}$ (see below), and
\begin{equation}
  k_{\rm H}=\omega-m\Omega_{\rm H}~~~{\rm with}~~~\Omega_{\rm H}=-\lim_{r\to r_{\rm h}}\frac{g_{t\varphi}}{g_{\varphi\varphi}}\,.
\label{eq:defkh}
\end{equation}
At the horizon and at infinity the couplings (between scalar and gravitational perturbations, and between perturbations
with different values of $l$) are subleading, and the perturbation equations can be written as two second-order
differential equations with the structure (with $Z^{lm}$ being either $\Phi^{lm}$ or $K^{lm}$)
\begin{align}
  Z^{lm}_{,r_*r_*}+k_H^2Z^{lm}&=O(r-r_{\rm h})~~~~~(r\to r_{\rm h})\nonumber\\
  \label{eq:expwave}
  Z^{lm}_{,r_*r_*}+\omega^2Z^{lm}&=O\left(\frac{1}{r^2}\right)~~~~~(r\to\infty)\,,
\end{align}
with boundary conditions 
\begin{align}
  Z^{lm}&\sim e^{-ik_{\rm H}r_*}~\,~~~{\rm at}~~(r\to r_{\rm h})\nonumber\\
\label{eq:bc}  Z^{lm}&\sim e^{\,i\omega r_*}~~~~~~~{\rm at}~~(r\to \infty)\,.
\end{align}
The tortoise coordinate $r_*(r)$ maps the region outside the BH horizon $r\in[r_{\rm h},\infty]$ into
$r_*\in[-\infty,+\infty]$; the function
\begin{equation}
\frac{dr}{dr_*}=F(r)
\end{equation}
behaves as $F(r)\sim r-r_{\rm h}$ for $r\to r_{\rm h}$, and $F(r)\to 1$ for $r\to\infty$. The explicit expression of
$F(r)$ can be found by requiring that the perturbation equations reduce, at the horizon and at infinity, to
Eq.~\eqref{eq:expwave}.

Actually, for rotating BHs in EdGB gravity there are different possible functions $F(r)$
satisfying this requirement.  In Paper I we have shown that by imposing a stronger requirement, i.e. that besides
Eq.~\eqref{eq:expwave},
\begin{equation}
  Z^{lm}_{,r_*r_*}+\omega^2Z^{lm}=\frac{l(l+1)}{r^2}Z^{lm}+O\left(\frac{1}{r^3}\right)\label{eq:expwave2}
\end{equation}
for $r\to\infty$, the function $F(r)$ is uniquely determined. To $O(\zeta^2)$, it is
\begin{align}
  F(r)=&\left(1-\frac{r_{\rm h}}{r}\right)\left\{1 -{\bar a}^2\frac{r_{\rm h}(r^2+rr_{\rm h}+r_{\rm h}^2)}{8r^3}\right.\nonumber\\
    &\left.-\zeta^2\left[\frac{r_{\rm h}}{3840 r^4}(147r^3+117r^2r_{\rm h}-526rr_{\rm h}^2+263r_{\rm h}^3)\right.\right.\nonumber\\
    &\left.\left.+{\bar a}^2\frac{r_{\rm h}}{30720 r^3}(375r^2+435rr_{\rm h}+343r_{\rm h}^2)
    \right]\right\}\nonumber\\
  &+O(\zeta^3)+O({\bar a}^3)\,.\label{eq:deftortoise}
\end{align}
The explicit expression of $F(r)$ to $O(\zeta^6)$ is given in the Supplemental Material~\cite{notebook}. We have
verified that with a different definition of the tortoise coordinate -- satisfying the condition~\eqref{eq:expwave} but
not the stronger condition~\eqref{eq:expwave2} -- the values of the QNMs are the same within the numerical error.

We remark that both the equations~\eqref{generalstructure-pol},~\eqref{generalstructure-ax} and the boundary conditions
for QNMs\,\eqref{eq:bc} are invariant for the transformation in $(\bar a,m)\to(-\bar a,-m)$ (note that $\Omega_{\rm
  H}\propto\bar a$), as long as axial perturbations change sign and polar perturbations remain the same.
Therefore, the solution and the quasi-normal modes frequencies are invariant for this transformation as well. This
implies that the $O(\bar a)$ corrections in the spin are odd in $m$, while the second-order corrections are even (see
e.g.\,\cite{1993ApJ...414..247K} and Paper I). Since the equations are quadratic at most in $m$, we shall make the
following ansatz for the QNM frequencies
\begin{align}
\omega=\omega_0+ {\bar a} \, m \omega_1+ {\bar a}^2 \,\left(\omega_{2a}+m^2 \omega_{2b}\right) + O({\bar a}^3)\,,
\label{eq:qnms-expansion}
\end{align}
where $\omega_{r}$ ($r=0,1,2a,2b$) do not depend on $m$. This will be confirmed by the actual QNM computation. 
%
\subsection{Perturbation equations}\label{sec:secondorder}
We shall here discuss the derivation and the numerical implementation of the perturbation equations to second order in
the spin. For the derivation in the non-rotating case and to first order in rotation, we refer the reader
to\,\cite{Salcedo2016} and to Paper I, respectively.

We  decompose the  metric  and scalar  field  perturbations  in terms  of  the perturbation  functions  of polar  parity
$\{H_0^{lm}(r),H_1^{lm}(r),H_2^{lm}(t,r),K^{lm}(r),\Phi^{lm}(r)\}$ and of  axial parity $\{h_0^{lm}(r),h_1^{lm}(r)\}$ as
in Eqs.~\eqref{eq:exp_scal}-\eqref{eq:exp_ax}.  The field equations~\eqref{eq:scalar}, \eqref{eq:metric},  linearized in
the perturbations and to second  order in the spin, yield a system of ordinary  differential equations, with the general
structure~\eqref{generalstructure-pol},~\eqref{generalstructure-ax}. We  remark that the equations  couple perturbations
with different parities, and with different values of the harmonic index $l$.

As discussed in~\cite{1993PThPh..90..977K,1993ApJ...414..247K,Pani2012scalar}, the couplings of the perturbations with
index $l$ to those with index $l\pm1$ can be neglected in the computation of the QNM spectrum to first order in the
spin. Similarly, as we are going to show, we can neglect the couplings between perturbations with index $l$
and with index $l\pm2$ in the computation of the QNMs to second order in the spin.

Let us consider the expansion in the spin of the perturbation functions with polar parities $Z_{\rm pol}^{lm}$
($=H_0^{lm}$, $H_1^{lm}$, ...) and with axial parities $Z_{\rm ax}^{lm}$ ($=h_0^{lm}$, $h_1^{lm}$):
\begin{equation}
  Z_{\rm pol/ax}^{lm} = Z_{\rm pol/ax}^{lm\,(0)}+ \bar{a}\, Z_{\rm pol/ax}^{lm\,(1)}
  + \bar{a}^2\, Z_{\rm pol/ax}^{lm\,(2)} \,.\label{eq:spinex}
\end{equation}
Since perturbations with index $l\pm1$ are always multiplied by $\bar a$ in the
equations~\eqref{generalstructure-pol},~\eqref{generalstructure-ax}, their second-order term in the
expansion~\eqref{eq:spinex} does not contribute to the equations; moreover, when they are multiplied by ${\bar a}^2$,
their first-order terms does not contribute as well. Similarly, perturbations with index $l\pm2$ contribute to the
equations~\eqref{generalstructure-pol},~\eqref{generalstructure-ax} with their $0$-order part in the
expansion~\eqref{eq:spinex} only.

Let us now assume that a source only excites a polar perturbation with a given harmonic index $l$. The rotation-induced
couplings in the field equations induce perturbations with axial parity and with harmonic index $l'\neq l$, but they
vanish in the non-rotating limit: $Z_{\rm ax}^{lm\,(0)}=Z_{\rm pol}^{l'm\,(0)}=0$. Axial parity perturbations with index
$l\pm1$ are excited through the rotation-induced couplings at first order in the spin, and are
$Z_{\rm ax}^{l\pm1m}=O(\bar a)$; similarly, polar parity perturbations with index $l\pm2$ are excited through the
rotation-induced couplings at second order in the spin, and are $Z_{\rm pol}^{l\pm2m}=O({\bar a}^2)$.  Therefore,
neglecting $O({\bar a}^3)$ terms, Eqs.~\eqref{generalstructure-pol}, \eqref{generalstructure-ax} reduce to
\begin{align}
  & {\cal{P}}_{l\,m}+ {\bar a} \,m\, \bar{\cal{P}}_{l\,m} + {\bar a}^2  \hat{\cal{P}}_{l\,m}+ \, m^2
  \, {\bar a}^2  {\bar{\bar{\cal{P}}}}_{l\,m} \notag \\
& +{\bar a}\left(Q_{l\,m} \tilde{\cal{A}}_{l-1\,m}+Q_{l+1\,m}\tilde{\cal{A}}_{l+1\,m}\right) = 0 \notag\\
 & {\cal{A}}_{l+1\,m}+ {\bar a} \,m\, \bar{\cal{A}}_{l+1\,m}  +{\bar a} \, Q_{l+1\,m} \tilde{\cal{P}}_{l\,m} \notag\\
&+ m \,{\bar a}^2\, Q_{l+1\,m} \check{\cal{P}}_{l\,m} = 0 \notag\\
 & {\cal{A}}_{l-1\,m}+ {\bar a} \,m\, \bar{\cal{A}}_{l-1\,m}  +{\bar a} \, Q_{l\,m} \tilde{\cal{P}}_{l\,m} \notag\\
&+ m \,{\bar a}^2\, Q_{l\,m} \check{\cal{P}}_{l\,m} = 0\,.
\label{polar-led-eqs}
\end{align}
These perturbations --~ on which we shall focus in this work --~ form a subset of the solutions of
Eqs.~\eqref{generalstructure-pol},~\eqref{generalstructure-ax}, called {\it polar-led sector}. 

A similar set of solutions is the {\it axial-led sector} of perturbations sourced by an axial perturbation with a given
$l$. We shall not consider axial-led perturbations, because they have no coupling between the metric and the scalar
field at zero-th order in rotation, and thus the QNMs are very close to those of GR\,\cite{Salcedo2016}.

The perturbation equations for the polar-led sector can be written as (see
Appendix \ref{app:eq})
\begin{widetext}
\begin{flalign}
  &A^{(I)}_{lm}+ \bar{a}^2 A^{(I)}_{2,lm}+ \bar{a}^2 \hat{A}^{(I)}_{lm}\left[Q_{lm}^2+Q_{l+1\,m}^2\right]+ \bar{a}^2
  \tilde{B}^{(I)}_{lm}\left[l Q^2_{l+1\,m}-(l+1)Q^2_{lm}\right] + i \  \bar{a} \ m C^{(I)}_{lm} + \bar{a}^2 m^2 E^{(I)}_{lm} && \notag\\ 
  & \ \ \ +Q_{lm} \bar{a} \left[\tilde{A}^{(I)}_{l-1\,m}+(l-1)B^{(I)}_{l-1\,m}\right] +Q_{l+1\,m} \bar{a}
  \left[\tilde{A}^{(I)}_{l+1\,m}-(l+2)B^{(I)}_{l+1\,m}\right] = 0\,, &&
\label{eq:eq1PL}
\end{flalign}

\begin{flalign}
  &l(l+1)\alpha^{(J)}_{lm}\!+\! l(l+1)\bar{a}^2 \alpha^{(J)}_{2,lm}\!-\! i m \ \bar{a} \left[\tilde{\beta}^{(J)}_{lm} +
    \zeta^{(J)}_{lm}\!-\!(l-1)(l+2)\xi^{(J)}_{lm}\right]+\bar{a}^2 \left[(l+1)(l-2)Q_{lm}^2+l(l+3)Q^2_{l+1\,m}
    \right]\hat{\alpha}^{(J)}_{lm} && \notag\\
  & \ \ \ + m^2 \bar{a}^2 \Delta^{(J)}_{lm} + \bar{a}^2 \left[l Q^2_{l+1\,m}-(l+1)Q_{lm}^2\right]\tilde{\eta}^{(J)}_{lm}+
  \bar{a}^2 \left[2 m^2 +Q^2_{lm}(l+1)(l^2-l+4)-Q^2_{l+1\,m} l (l^2+3l+6)\right]\tilde{\gamma}^{(J)}_{l} && \notag\\
  & \ \ \  + \bar{a} Q_{lm} (l+1)\Big\{(l-1)\tilde{\alpha}^{(J)}_{l-1\,m} -
  \eta^{(J)}_{l-1\,m}+(l-2)(l-1)\gamma^{(J)}_{l-1\,m} \Big\} &&\notag\\
  & \ \ \ +\bar{a} Q_{ {l} +1\,m} l\Big\{(l+2)\tilde{\alpha}^{(J)}_{l+1\,m} +  \eta^{(J)}_{l+1\,m}-
  (l+2)(l+3)\gamma^{(J)}_{l+1\,m} \Big\} =0\,, &&
\label{eq:eq2PL}
\end{flalign}

\begin{flalign}
  & (l+1)(l+2)\beta^{(J)}_{l+1\,m} + i m \ \bar{a}\left[\tilde{\alpha}^{(J)}_{l+1\,m} + \eta^{(J)}_{l+1\,m} +
    l(l+3)\gamma^{(J)}_{l+1\,m}\right] &&\notag\\
  &\ \ \ + \bar{a} Q_{l+1\,m}(l+2) \Big\{l\tilde{\beta}^{(J)}_{lm} -
  \zeta^{(J)}_{lm}-(l-1)l \xi^{(J)}_{lm} \Big\} && \notag\\
  & \ \ \ + i m \ \bar{a}^2  Q_{l+1\,m} \left[2 \hat{\alpha}^{(J)}_{lm}-(l+2)\Delta^{(J)}_{lm}+
    \tilde{\eta}^{(J)}_{lm}+(l-1)(l+4)\tilde{\gamma}^{(J)}_{lm}\right] = 0 \,,&&
\label{eq:eq3PL}
\end{flalign}

\begin{flalign}
  & l(l-1)\beta^{(J)}_{l-1\,m} + i m \ \bar{a}\left[\tilde{\alpha}^{(J)}_{l-1\,m} + \eta^{(J)}_{l-1\,m} +
    (l-2)(l+1)\gamma^{(J)}_{l-1\,m}\right] &&\notag\\
  &\ \ \ + \bar{a} Q_{lm}(l-1) \Big\{(l+1)\tilde{\beta}^{(J)}_{lm} +
  \zeta^{(J)}_{lm}+ (l+1)(l+2)\xi^{(J)}_{lm} \Big\} && \notag\\
  & \ \ \ + i m \ \bar{a}^2 Q_{lm} \left[2 \hat{\alpha}^{(J)}_{lm}+(l-1)\Delta^{(J)}_{lm}+ \tilde{\eta}^{(J)}_{lm}
    +(l-3)(l+2)\tilde{\gamma}^{(J)}_{lm}\right] = 0 \,,&&
\label{eq:eq3bPL}
\end{flalign}

\begin{flalign}
  & l(l-1)(l+1)(l+2) (s_{lm}+\bar{a}^2 s_{2,lm}) - i m \ \bar{a} (l-1)(l+2)f_{lm} + \bar{a}^2 \Big[2 m^2 + Q_{lm}^2
    (l+1)(l^2-l+4) && \notag\\
    & \ \ \ -Q^2_{l+1\,m} l(l^2+3 l+6) \Big]\tilde{g}_{lm} + \bar{a}^2 \left[2 m^2 -l(l+1) +(l+1)(l+2) Q_{lm}^2 +
    l(l-1) Q^2_{l+1\,m}\right]\hat{k}_{lm} && \notag\\
  & \ \ \ + \bar{a}^2 \Big\{8 m^2-2 l(l+1)-Q_{lm} l^2 (l+1)\left[4(l-2)-l(l+1)(l+4)\right]-Q_{l+1\,m}^2 l
  \left[4(l+3)-l(l+1)(l-3)\right] \Big\}\hat{s}_{lm} && \notag\\
  & \ \ \ - Q_{lm}\Big\{\bar{a} (l-1)(l+1)(l+2) g_{l-1\,m}\Big\}  +  Q_{l+1\,m}\Big\{\bar{a}l(l-1)(l+2)g_{l+1\,m}
  \Big\} = 0\,, &&
\label{eq:eq4PL}
\end{flalign}

\begin{flalign}
  & l(l+1)(l+2)(l+3) t_{l+1\,m} + i m \ \bar{a} \  l(l+3)g_{l+1\,m}  - \bar{a} \ Q_{l+1\,m}\Big\{l(l+2)(l+3)
  f_{lm}\Big\} && \notag\\
  & \ \ \ + i m \ \bar{a}^2 Q_{l+1\,m}\left[(l-2)(l+3)\tilde{g}_{lm}- 2(l+3) \hat{k}_{lm} + 4 (l-1)(l+3)
    \hat{s}_{lm}\right] = 0 \,,&&
\label{eq:eq5PL}
\end{flalign}

\begin{flalign}
  & (l-1)(l-2)l(l+1) t_{l-1\,m} + i m \ \bar{a}(l-2)(l+1)g_{l-1\,m}   + \bar{a} \ Q_{lm}\Big\{(l-1)(l-2)(l+1)f_{lm}
  \Big\} && \notag\\
  & \ \ \ + i m \ \bar{a}^2 Q_{lm} \left[(l-2)(l+3) \tilde{g}_{lm}+2 (l-2) \hat{k}_{lm} +4 (l-2)(l+2)\hat{s}_{lm}
    \right] = 0\,, &&
\label{eq:eq5bPL}
\end{flalign}
\end{widetext}
where the quantities $A^{(I)}$, $C^{(I)}$, $\alpha^{(J)}$, $\beta^{(J)}$, etc. ($I=0,\dots,4$, $J=0,1$) are combinations
of the perturbation functions and of their derivatives.  We have followed and expanded the notation
of~\cite{Kojima1992}; at variance with~\cite{Kojima1992}, the dependence on the spin $\bar a$ has been factored out;
therefore, the quantities appearing in Eqs.~\eqref{eq:eq1PL}-\eqref{eq:eq5bPL} depend on $\zeta$ but not on $\bar
a$. Moreover, we have introduced new quantities ($A_{2,lm}$, $E^{(I)}_{lm}$, ${\hat\alpha}^{(J)}_{lm}$,
$\Delta^{J}_{lm}$, ${\tilde\eta}^{(J)}_{lm}$, ${\tilde\gamma}^{(J)}_{lm}$, $s_{2,lm}$, ${\tilde g}_{lm}$, ${\hat
  k}_{lm}$, ${\hat s}_{lm}$), which appear at second order in the spin.
The explicit expressions of the quantities in Eqs.~\eqref{eq:eq1PL}-\eqref{eq:eq5bPL}, up to $O(\zeta^6)$, are given in
the Supplemental Material~\cite{notebook}.

We remark that since some of the tensor spherical harmonics identically vanish for $l=0,1$, it is possible to exploit
the residual gauge freedom to set to zero the axial perturbations (see e.g.\,
\cite{Blazquez-Salcedo:2017txk}). Therefore, Eqs.~\eqref{eq:eq1PL}-\eqref{eq:eq5bPL} are valid for $l\ge2$ and, in the
case $l=2$ (in which polar perturbations with index $l$ are coupled with axial perturbations with index $l\pm1$), the
axial perturbations with index $l-1$ can be set to zero.

With appropriate combinations of the perturbation equations, we can find $H^{lm}_0$ and $H_2^{lm}$ as algebraic expressions in
terms of $H_1^{lm}$ and $K^{lm}$. Thus, calling $\xi^{lm}=\frac{d}{dr}\Phi^{lm}$ and defining 
\begin{equation}
\bm \Psi^{lm} = 
\begin{pmatrix}
H_1^{lm}\\
K^{lm}\\
\Phi^{lm}\\
\xi^{lm}\\
h_0^{l+1\,m}\\
h_1^{l+1\,m}\\
h_0^{l-1\,m}\\
h_1^{l-1\,m}\\
\end{pmatrix}
\label{eq:psi_edgb-2}
\end{equation}
we can cast our equations (for given values of $l,m$) as
\begin{equation}
\frac{d}{dr} \bm{\Psi}_{lm} + \hat{P}_{lm}\bm{\Psi}_{lm} = 0
\label{eq:vect-system-2nd}
\end{equation}
where $\hat{P}_{lm} = \hat{P}^{(0)}_{lm} + \bar{a} \hat{P}^{(1)}_{lm} + \bar{a}^2 \hat{P}^{(2)}_{lm}$ is an
eight-dimensional square matrix.
In the $l=2$ case, since axial perturbations with $l=1$ can be set to zero,
$\bm \Psi^{2m}= \{H_1^{2m}, K^{2m},\Phi^{2m},\xi^{2m}, h_0^{3\,m}, h_1^{3\,m} \}$
and the matrix $\hat{P}_{lm}$ is six-dimensional.

As discussed in Sec.~\ref{sec:qnm}, the perturbation functions behave at the horizon and at infinity as in
Eq.~\eqref{eq:bc0}.  The QNMs are the perturbations satisfying ingoing boundary conditions at the horizon
($\sim e^{-i k_H r_*}$) with $k_H$ given in Eq.~\eqref{eq:defkh}, and outgoing boundary conditions at infinity
($\sim e^{i \omega r_*}$). To find the QNM (complex) frequencies, we follow the same approach as in Paper I (see also
e.g.~\cite{Ferrari:2007rc,AdvancedMethods}): we define an eight-dimensional (six-dimensional for $l=2$) square matrix
whose columns are four (three) independent solutions satisfying the QNM boundary conditions at the horizon (superscript
$^{(-)}$), and four (three) independent solutions satisfying the boundary conditions at infinity (superscript $^{(+)}$),
evaluated at a matching point $r_m$: for $l=2$, we can write
\begin{equation}
{X}=
\begin{pmatrix}
\bm \Psi_{1a}^- & \bm \Psi_{1b}^- & \bm \Psi_{1c}^- & \bm \Psi_{1a}^+ & \bm \Psi_{1b}^+ & \bm \Psi_{1c}^+\\
\end{pmatrix}\,.
\label{eq:X_edgb_2}
\end{equation}
The QNMs are found by imposing the condition
\begin{equation}
{\rm det} {X}(\omega^{nlm})=0\,.
\label{eq:detX2bis}
\end{equation}
As discussed in Paper I (see also~\cite{Salcedo2016}), the gravitational QNMs of black holes in EdGB gravity belong to
two classes: {\it gravitational-led} modes (which reduce to the gravitational QNMs of GR as $\zeta=0$) and {\it
  scalar-led} modes (which reduce to the scalar QNMs of GR as $\zeta=0$). In this article we only consider
gravitational-led modes, which are expected to be excited with larger amplitudes by realistic
sources\,\cite{Barausse:2014tra,Salcedo2016}.
\subsection{Spin expansion of the quasi-normal modes}\label{sec:spinexp}

\subsubsection{Taylor expansion}
As discussed in Sec.~\eqref{sec:qnm}, the QNM frequencies at second order in the spin (see
Eq.~(\ref{eq:qnms-expansion})) can be written as
\begin{align}
  \omega^{nlm}({\bar a},\zeta)&=\omega^{nl}_0 (\zeta)+ {\bar a}\,m \omega^{nl}_1 (\zeta)\nonumber\\
  &+ {\bar a}^2 \left[\omega^{nl}_{2a}(\zeta)+\,m^2 \omega^{nl}_{2b} (\zeta)\right] +\mathcal{O}({\bar a}^3)
\label{eq:qnms-expansion-2}
\end{align}
where $\omega_0^{nl}(\zeta)$ is the QNM frequency in the static case. Eq.~\eqref{eq:qnms-expansion-2} is a Taylor
expansion around $\bar a=0$. Therefore, once the function $\omega^{nlm}({\bar a},\zeta)$ is found from the numerical
solution of the equation, its derivatives with respect to $\bar a$ yield the functions $\omega^{nl}_{r}(\zeta)$
($r=0,1,2a,2b$). The separation between $\omega_{2a}$ and $\omega_{2b}$ is obtained by repeating the computation for
different values of $m$.
\subsubsection{Pad\'e resummation}
The Taylor expansions\,\eqref{eq:qnms-expansion-2} can be resummed using {\it Pad\'e approximants}
(see~\cite{Damour:1997ub,press2007numerical}). The Pad\'e resummation, which replaces polynomials with rational functions,
often improves the convergence of an expansion. This technique has been applied, for instance, to post-Newtonian
expansions~\cite{Damour:1997ub}, and more recently in the computation of BH sensitivities in EdGB
gravity~\cite{Julie:2019sab,Julie:2022huo}. Pad\'e resummation also improves the convergence of the spin expansion of BH
QNMs\,\cite{Hatsuda:2020egs}, as we shall discuss in Sec.~\ref{subsec:truncation}.

Given a Taylor expansion $T_K(x)$ of order $K$ around $x=0$, we can construct a Pad\'e approximant $P[M,N]$, with $M,N$
integer numbers such thar $M+N=K$, given by
\begin{align}
P[M,N] (x)= \frac{A_0+ A_1 x + A_2 x^2 + ... + A_M x^M}{B_0+ B_1 x + B_2 x^2 + ... + B_N x^N}
\label{eq: pade-general}
\end{align}
such that $P[M,N](x) = T_K(x)$ up to order $K$. Solving order by order in $x$, the coefficients $A_0,... A_M$, $B_0,
..., B_N$ can be determined as combinations of the Taylor expansion coefficients.

Since the Taylor expansion to second order is not accurate for QNMs of rotating BHs with large spins (we remind that a
BH in the aftermath of a binary coalescence has tipically $\bar a\sim0.7)$, we shall perform a Pad\'e resummation of the
second-order expansion~\eqref{eq:qnms-expansion-2}. In this case the Taylor approximant of $\omega^{nlm}({\bar
  a},\zeta)$ is of second order, and (for each $\zeta,n,l,m\neq0$) the possible choices of Pad\'e approximants are:
\begin{align}
  &P[1,1](\bar{a},\zeta) =\nonumber \\
  &\frac{m \,\omega_0^{nl}(\zeta) \omega_1^{nl}(\zeta) + \left[m^2 \,{\omega_1^{nl}}^2(\zeta)-
      \omega_0^{nl}(\zeta) \omega_2^{nlm}(\zeta)\right]\bar{a}}{m\,\omega_1^{nl}(\zeta)-\omega_2^{nlm}(\zeta) \bar{a}}
\label{eq: pade-11-coefficients}
\end{align}
and
\begin{align}
  &P[0,2](\bar{a},\zeta)= \nonumber \\
  &\frac{{\omega_0^{nl}}^3(\zeta)}{{\omega_0^{nl}}^2(\zeta)+ \bar{a}^2 m^2  {\omega_1^{nl}}^2(\zeta)- \bar{a}
    \omega_0^{nl}(\zeta) \left[m \omega_1^{nl}(\zeta)
    + \bar{a} \omega_{2}^{nlm}(\zeta)\right]} 
\label{eq: pade-02-coefficients}
\end{align}
where we remind that $\omega_2^{nlm}=\omega_{2a}^{nl}+m^2\omega_{2b}^{nl}$. Note that since the QNMs are complex, the
coefficients of the Taylor and Pad\'e approximants are complex as well.

As suggested in~\cite{Damour:1997ub}, we shall use the ``diagonal'' Pad\'e, $P[1,1]$, unless it is not accurate due to the
presence of a pole or a reduction of order in the polynomials, in which case we instead use $P[0,2]$. In practice, for
the QNMs with $n=0$, $l=2,3$ we shall always use $P[1,1]$ except for $m=0$ (since Eq.~\eqref{eq: pade-11-coefficients}
reduces to a constant) and for the imaginary parts of the modes with $m=\pm1$, for which $P[1,1]$ has a pole close to
the spin interval which we have considered. A similar computation has been done in\,\cite{Hatsuda:2020egs}, where
$P[1,1]$ was used for all values of $m$.

\section{Results}\label{sec:results}

By performing the numerical integration explained in the previous section, we find the functions
$\omega^{nl}_{r}(\zeta)$, where $r =0, 1, 2a, 2b$ (see Eq.~\eqref{eq:qnms-expansion-2}). As discussed in
Sec.\,\ref{sec:secondorder}, we focus on gravitational-led modes in the polar-led sector. We have computed the
fundamental (i.e., $n=0$) QNMs with $l=2,3$. We have not considered $n>0$ QNMs because our direct-integration approach
it not accurate in the computation of overtones\,\cite{chandrasekhar1975quasi}, and thus is not possible to extract the
EdGB correction for those modes.
%
\subsection{Estimates of the truncation errors}\label{subsec:truncation}
In this work, we have expanded both the background (Sec.~\ref{subsec:eqrot}) and the perturbation equations
(Sec.~\ref{sec:secondorder}) in the spin $\bar a$, up to second order, and in the dimensionless coupling constant
$\zeta$, up to sixth order. The QNMs have then been expanded in the spin to second order, and resummed using Pad\'e
approximants.
\subsubsection*{Expansion in the spin}
In order to assess the accuracy of the expansion in the spin, we have considered the slow-rotation expansion for
rotating BHs in GR\,(a similar approach has been followed in\,\cite{Wagle:2021tam}). We have computed the QNMs within
the slow rotation approximation, firstly to first order (neglecting $O(\bar a^2)$ terms in the background and in the
perturbation equations) and then to second order; the QNMs have then been Taylor-expended to the same order. Moreover,
the QNMs at second order in the spin have been resummed using Pad\'e approximants. Finally, we have compared the
frequencies of these modes with those of Kerr BHs (see e.g.~\cite{bertiweb}), by computing the discrepancies
\begin{equation}
  \delta \omega^{nlm}(\bar a) = \frac{\omega^{nlm}_{{\rm T,P}}-\omega^{nlm}_{\rm Kerr}}{\omega^{nlm}_{\rm Kerr}}\,
  \label{eq:discr}
\end{equation}
where the subscripts 'T' and 'P' refer to the modes (computed in slow-rotation expansion) Taylor-expanded and Pad\'e
resummed, respectively, while the subscript 'Kerr' refers to the modes of Kerr BHs.
\begin{figure}
	\includegraphics[width=8cm]{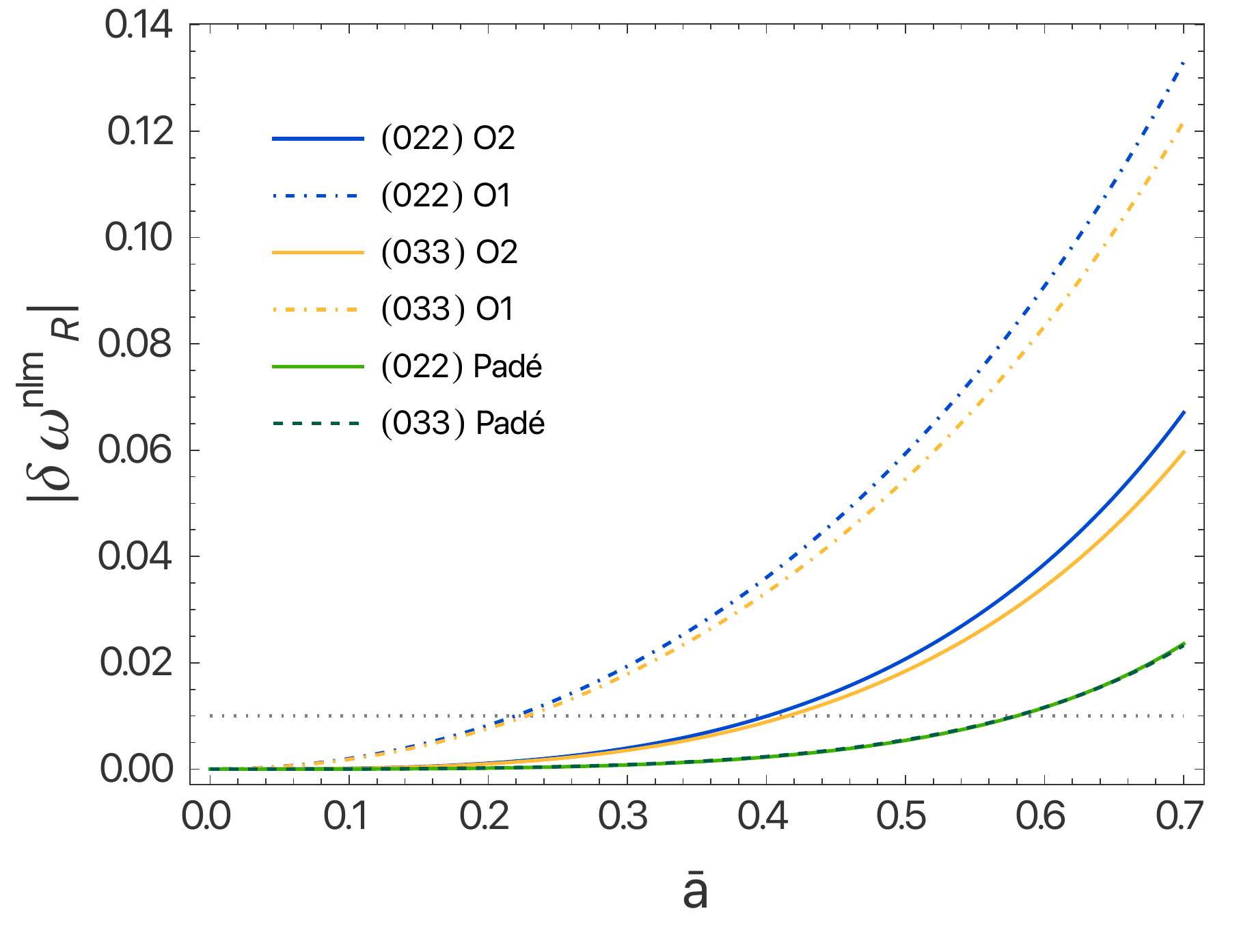}
	\includegraphics[width=8cm]{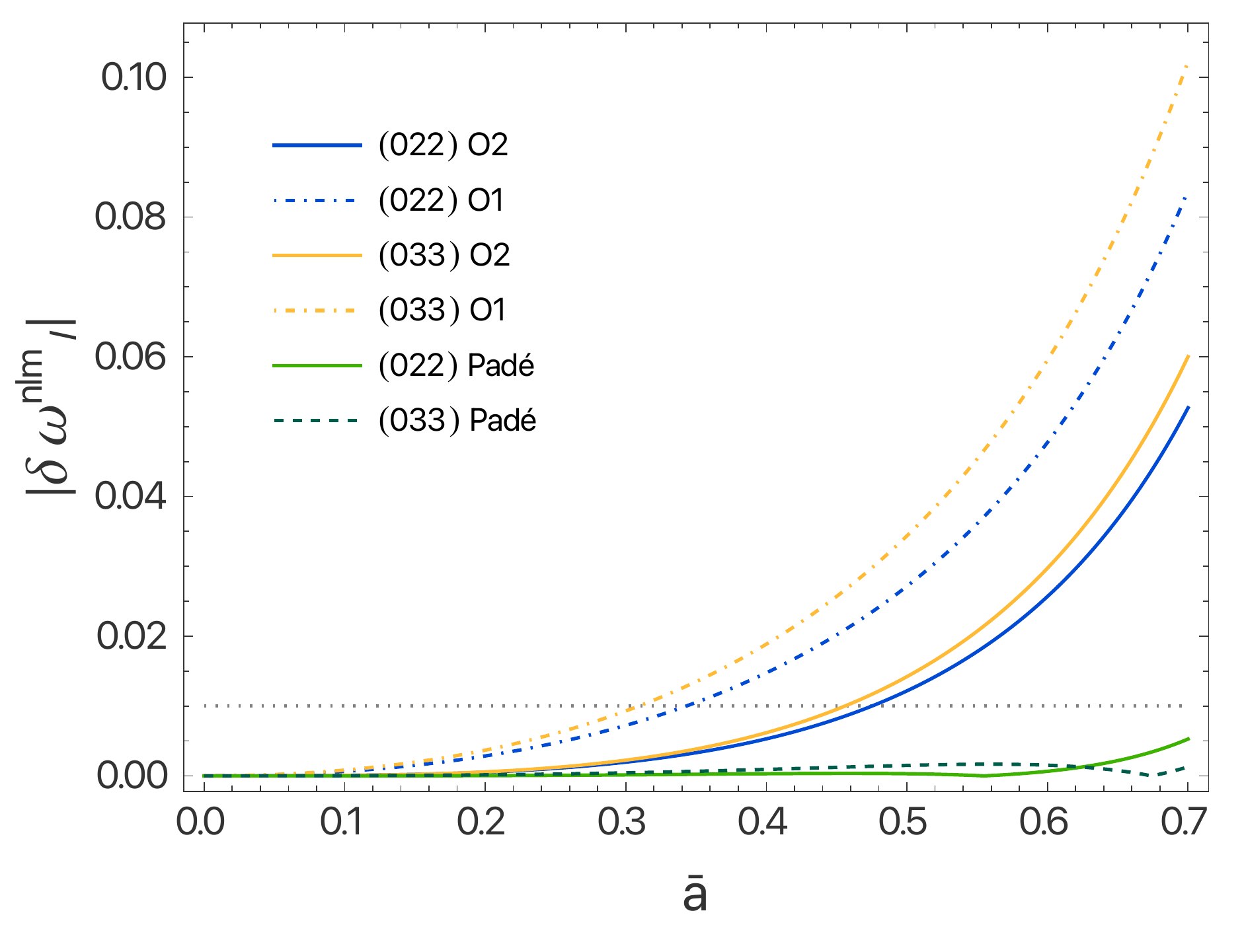}
	\caption{Real (upper panel) and imaginary (lower panel) parts of the relative difference between the QNMs of
          Kerr BHs and those of rotating BHs computed within the slow-rotation approximation, for the
          $(nlm)=(022),\,(033)$ modes. The slow-rotation expansion is performed to first order ($O1$), to second order
          ($O2$), to second order with Pad\'e resummation ({\it Pad\'e}). The horizontal dotted line represents a $1\%$
          error.}
	\label{fig:delta-kerr-SR}
\end{figure}
\begin{figure*}[hpt]
	\includegraphics[width=8cm]{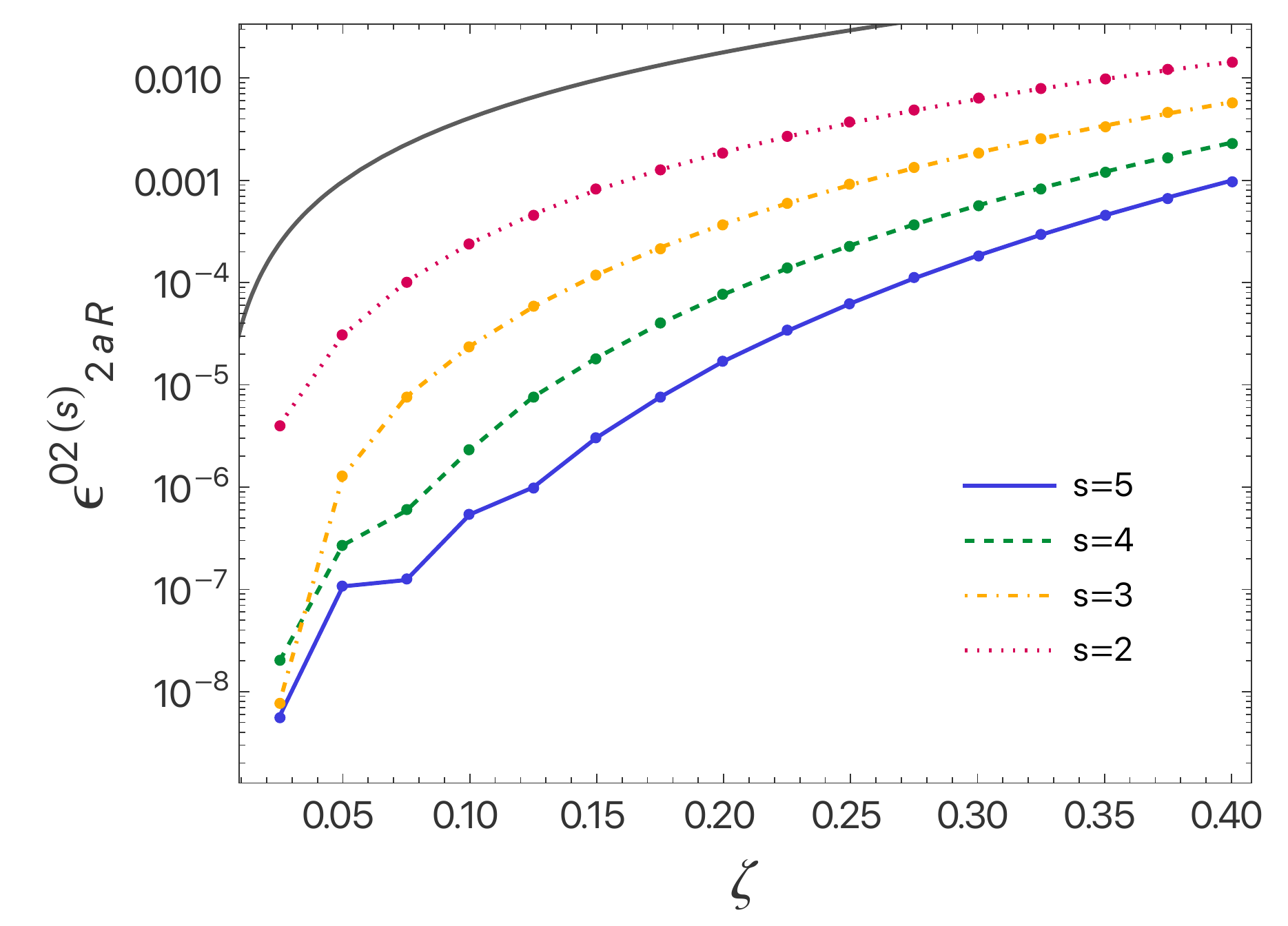}
	\includegraphics[width=8cm]{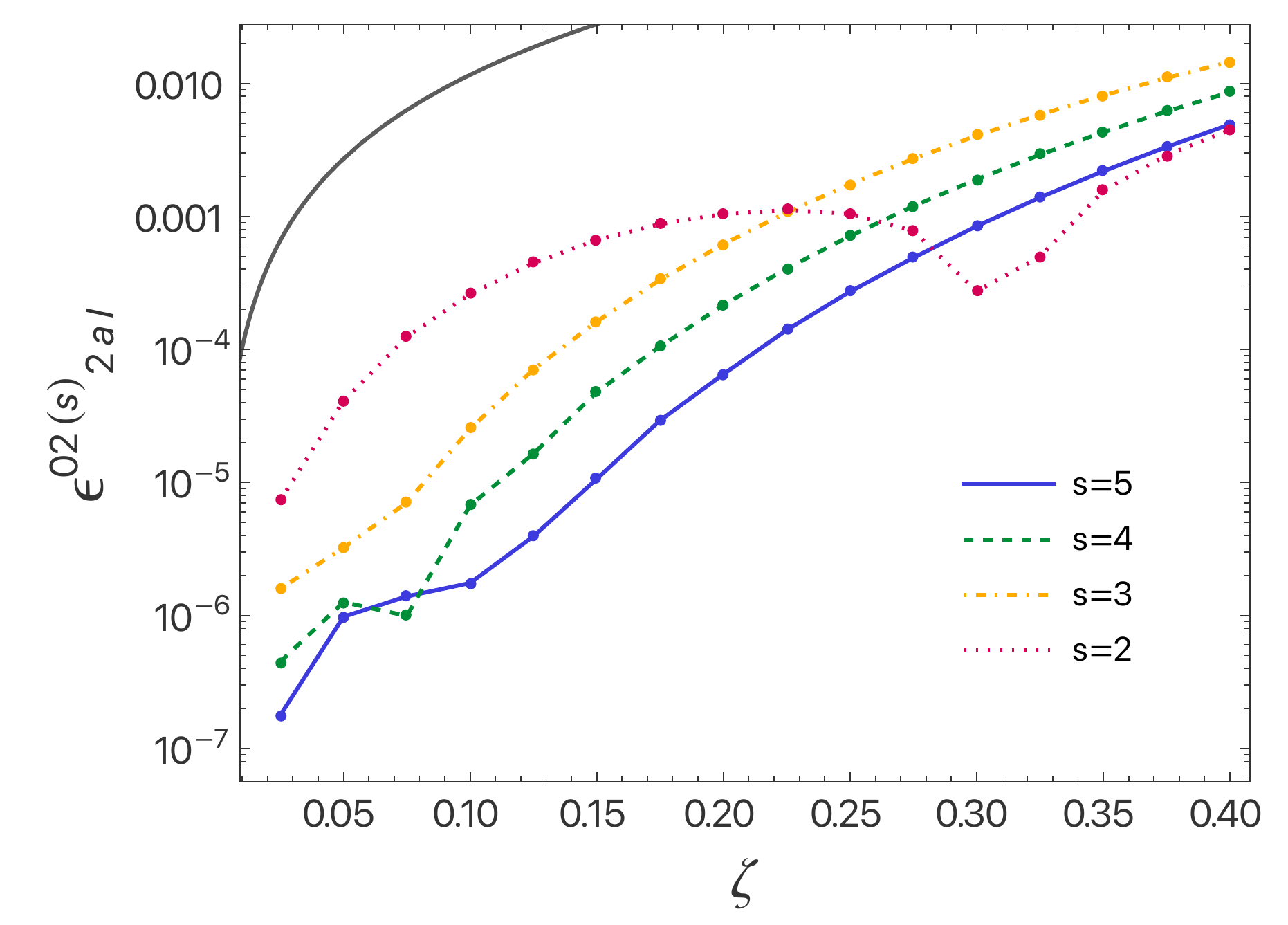}
	\includegraphics[width=8cm]{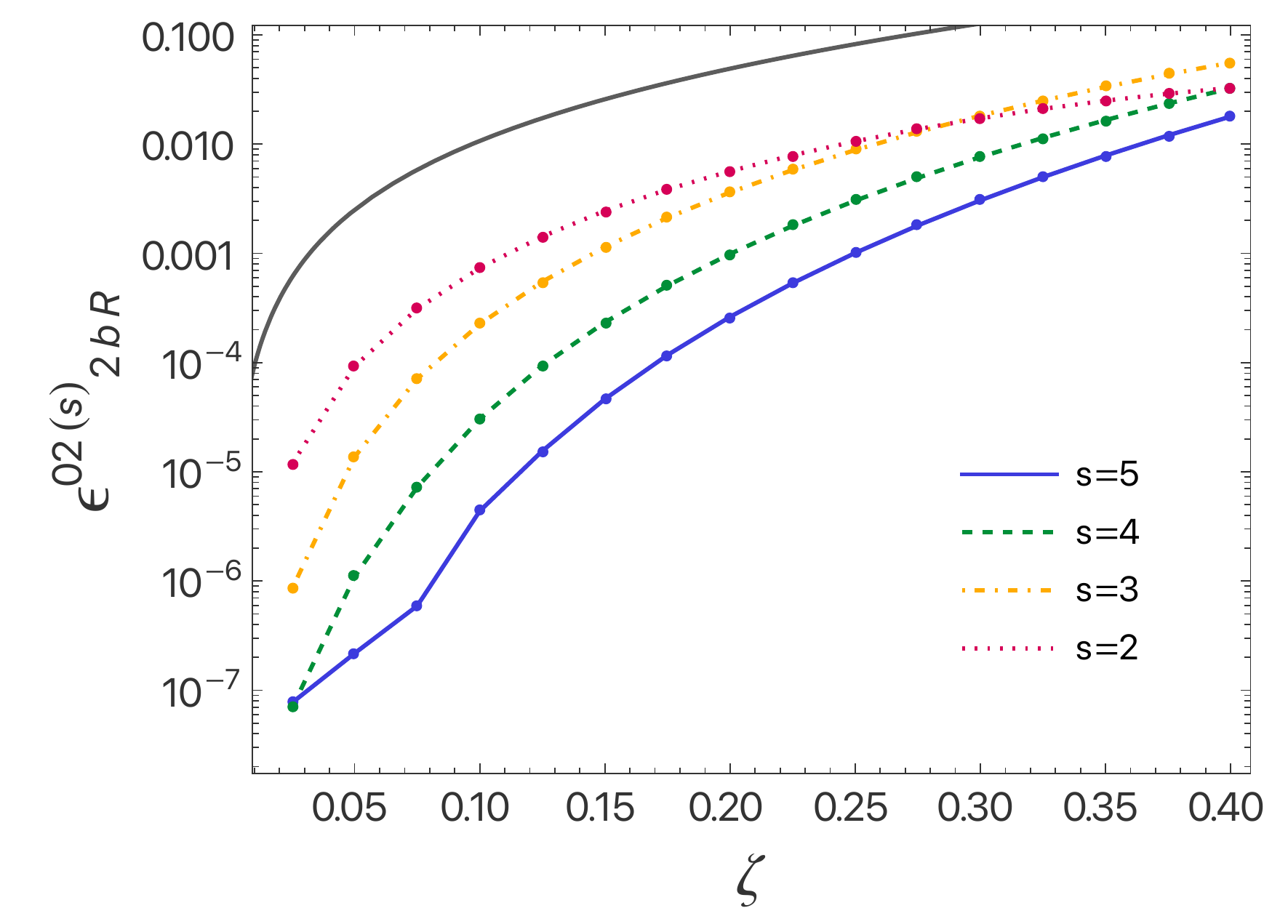}
	\includegraphics[width=8cm]{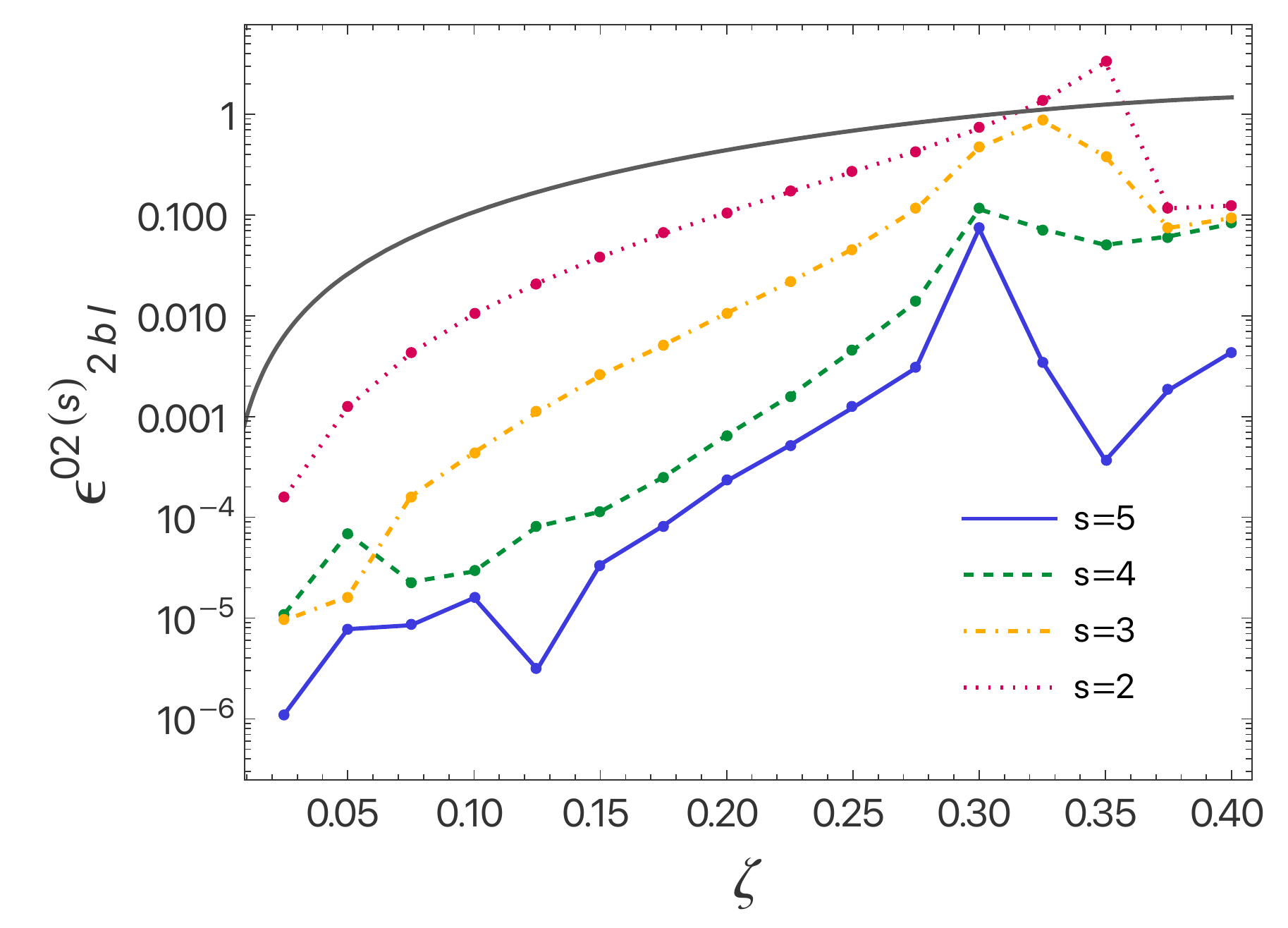}
	\caption{Truncation errors at order $s\le5$ for $\omega_{2a}(\zeta)$ (upper panels) and $\omega_{2b}(\zeta)$
          (lower panels), for the $(nl)=(02)$ QNM. The real and imaginary parts are shown in left panels and right
          panels, respectively. The truncation error is always smaller than the relative correction due to gravity
          modifications (solid curves).}\label{fig:truncerr-w2}
\end{figure*}

In Fig.~\ref{fig:delta-kerr-SR} we show real and imaginary parts of the discrepancies~\eqref{eq:discr} as functions of
$\bar a$, for the QNMs with $(nlm)=(022)$ and $(nlm)=(033)$, which are expected to be the most excited in typical binary
BH coalescences\,\cite{Bhagwat:2016ntk,Yang:2017zxs,Cabero:2019zyt,Ghosh:2021mrv}. The curves labeled $O1$, $O2$ show
the discrepancies between the modes of Kerr BHs and those computed within the slow-rotation approximation, to $O(\bar
a)$ and to $O({\bar a}^2)$, respectively. The curves labeled {\it Pad\'e} show the discrepancies with the $O({\bar
  a}^2)$ modes resummed using Pad\'e approximants (see Sec.~\ref{sec:spinexp}).  We see that at first order, the
discrepancy of the Taylor expansion is smaller than $1\%$ as long as $\bar a \lesssim 0.22$. Including the second order
correction, the discrepancy is smaller than $1\%$ for
$\bar a\lesssim 0.4$. The Pad\'e resummation improves the accuracy of the expansion, which is accurate to $\sim1\%$ for
$\bar a\lesssim0.6$ and to $\sim2\%$ for $\bar a\lesssim0.7$.

An analysis of the modes with different values of $m$ shows the same (or better) accuracy for the Pad\'e-resummed modes,
but we need to employ the approximant $P[0,2]$ instead of $P[1,1]$ in two cases: the modes with $m=0$ (for which
Eq.~\eqref{eq: pade-11-coefficients} reduces to a constant) and the imaginary parts of the modes with
$m=\pm1$.\footnote{In the latter case, the Pad\'e approximant $P[1,1]$ leads to a larger error, compared with that of
  the Taylor approximant, for $\bar a\sim0.7$; we think this is due to the presence of a pole close to the considered
  range of values for the spin. If, instead, we use $P[0,2]$ for the imaginary parts of the modes with $n=0$, $l=2,3$,
  $m=\pm1$, the error is smaller than $1\%$ for $\bar a\lesssim0.7$.}

These results (which are similar to those found in\,\cite{Hatsuda:2020egs}, with the exception of those for which we have used the $P[0,2]$ approximant, finding better accuracy) provide an indication that a
second-order computation of QNMs may be accurate for $\bar a\lesssim0.4$ ($\bar a\lesssim0.7$ with Pad\'e resummation)
for EdGB gravity as well.  In the following, then, we shall mostly consider values of the spin in the range $\bar
a\in[0,0.7]$.
\begin{figure*}[hpt]
	\includegraphics[width=8cm]{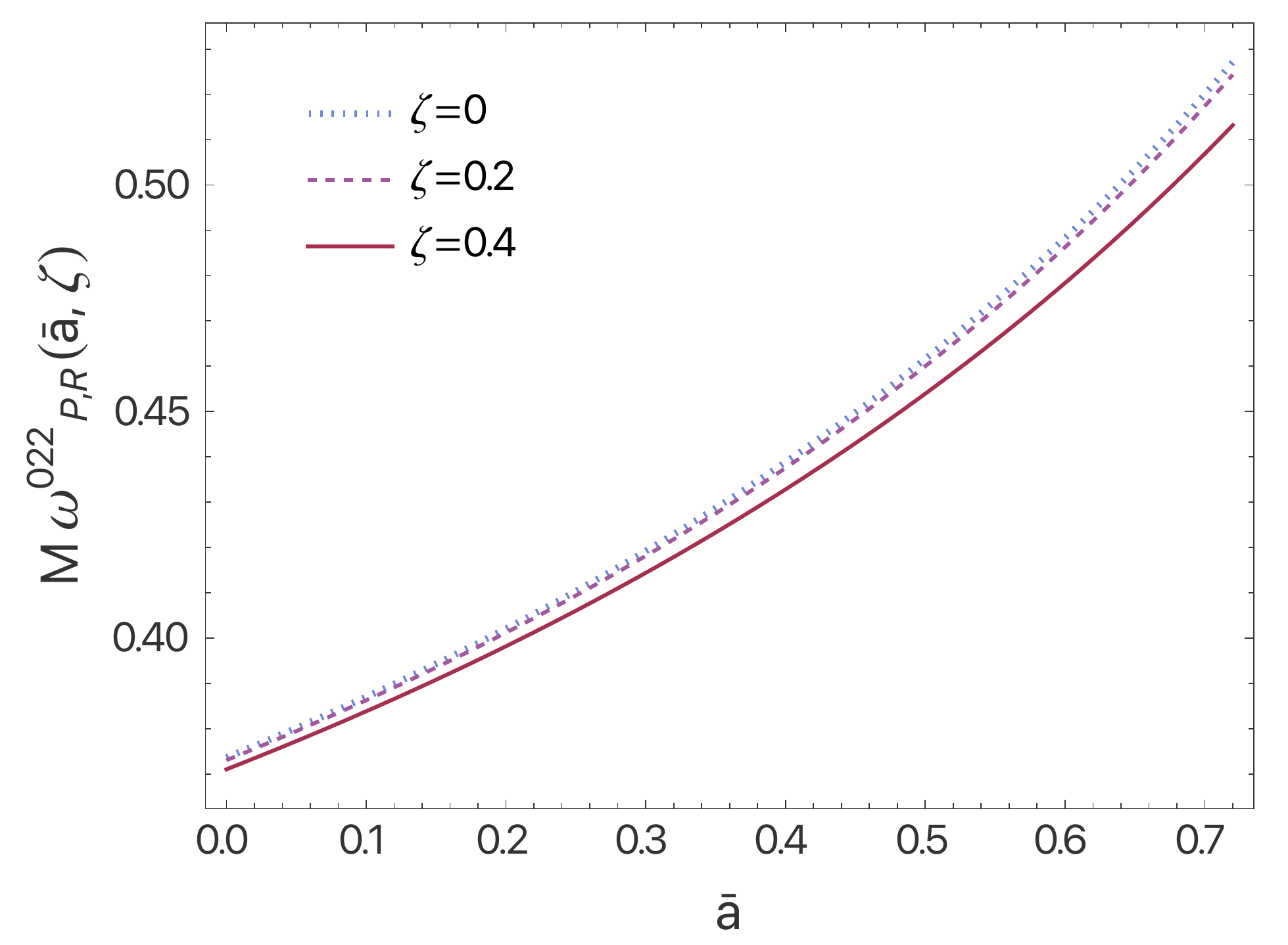}
	\includegraphics[width=8cm]{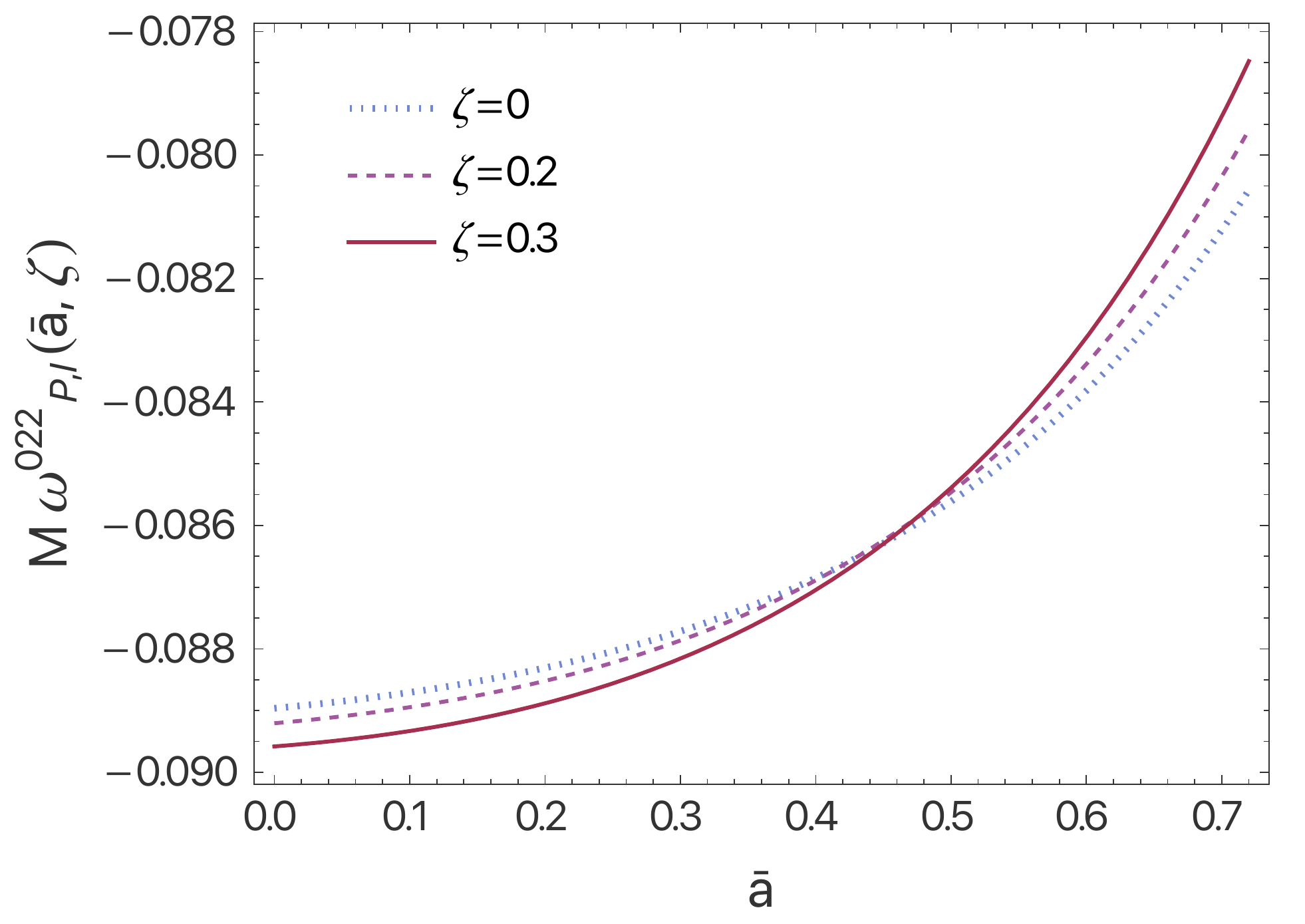}
	\includegraphics[width=8cm]{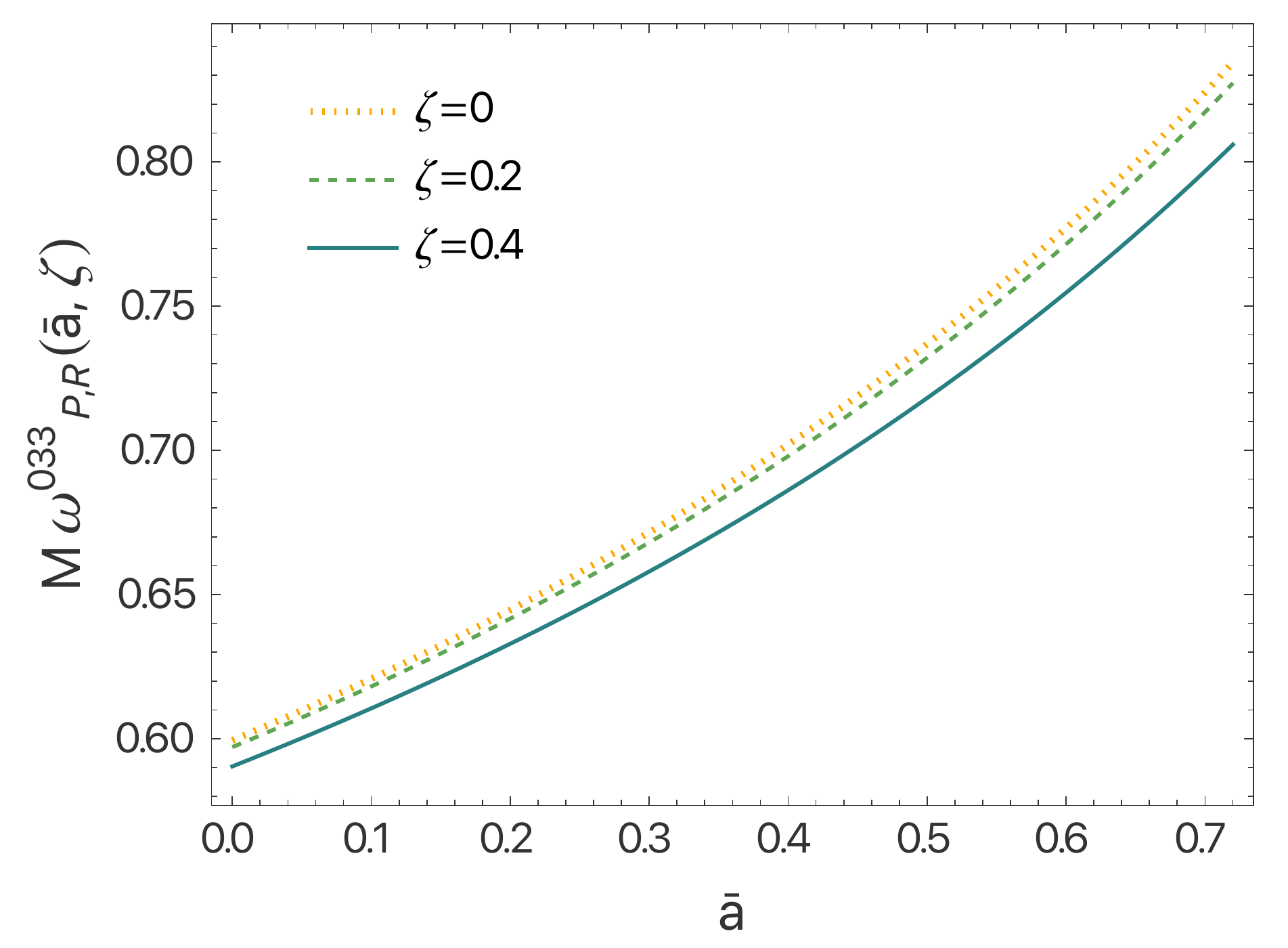}
	\includegraphics[width=8cm]{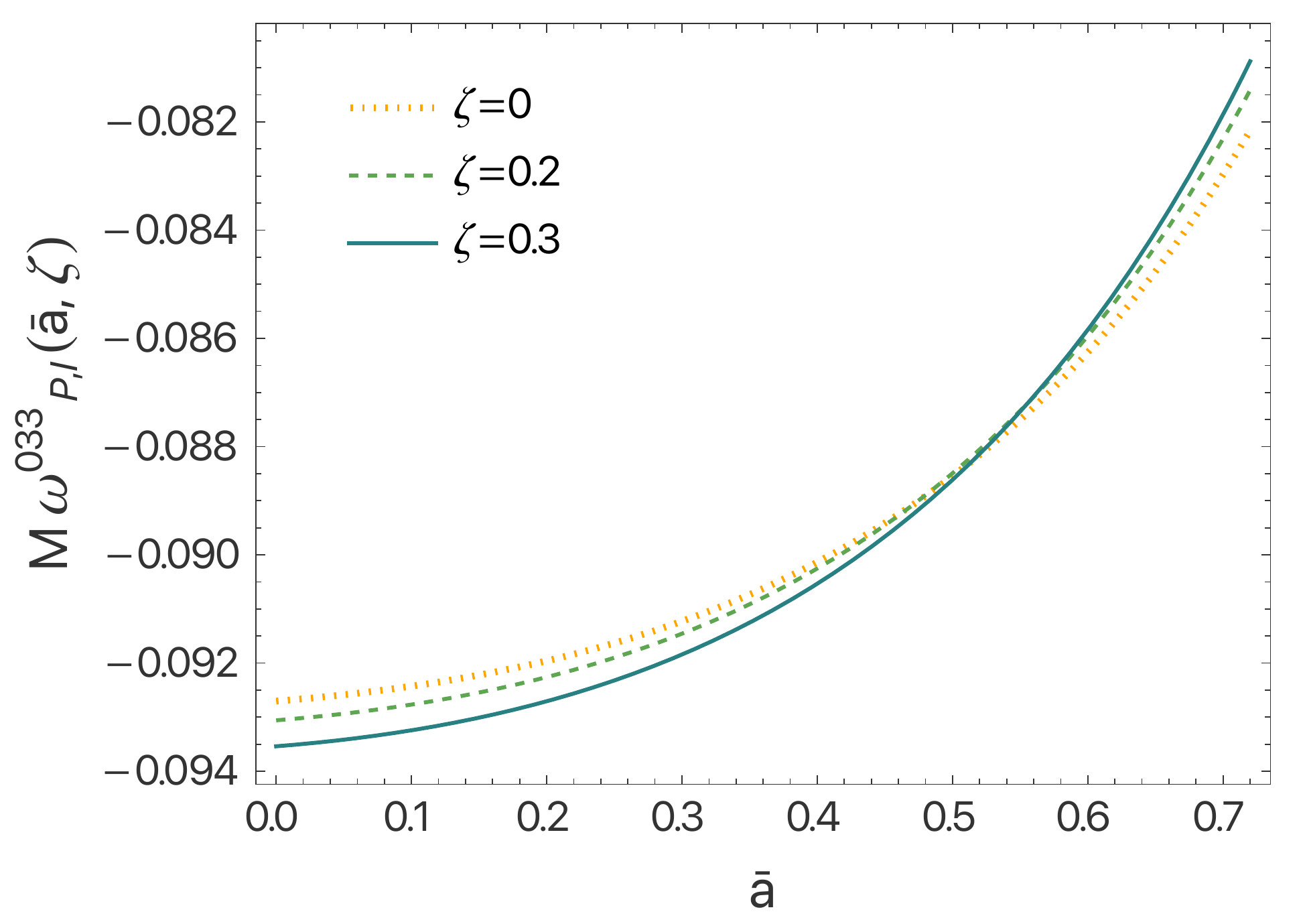}
	\caption{Real (upper panels) and imaginary (lower panels) parts of the $(nlm)=(022)$ (left panels) and
          $(nlm)=(033)$ (right panels) QNMs, evaluated using Pad\'e approximants, as functions of the spin, for different
          values of $\zeta$.}
	\label{fig:wtot}
\end{figure*}
\subsubsection*{Expansion in the coupling constant}
To assess the accuracy of the expansion in the dimensionless coupling $\zeta$, we have computed the functions
$\omega_r^{nl(s)}(\zeta)$\,\eqref{eq:qnms-expansion-2}, by expanding the background and the perturbation equations up to
order $s$ in $\zeta$ and up to second order in $\bar a$; we have repeated the computation for $s=2,\dots,6$, denoting
the functions computed in this way as $\omega_r^{nl(s)}(\zeta)$. We then define the truncation error at order $s$ of
$\omega_r^{nl}(\zeta)$ as:
\begin{align}
  \epsilon^{nl(s)}_{r \, R,I} (\zeta) = \frac{\left|\omega_{r \, R,I}^{nl(s+1)}(\zeta)-
    \omega_{r \, R,I}^{(s)}(\zeta)\right|}{\left|\omega_{r \, R,I}^{nl(s)}(\zeta)\right|}
\label{eq:truncerr}
\end{align}
where $r=0,1, 2a, 2b$ and the subscripts $R,I$ refer to the real and imaginary parts of the complex frequencies.

This analysis has been performed in Paper I to first order in the spin, i.e. for $r=0,1$. We here extend this
computation to the the discrepancies of the $O({\bar a}^2)$ contributions, i.e. for $r=2a$, $2b$. The truncation errors
for these functions are shown in Fig.~\ref{fig:truncerr-w2}. We see that (as for $r=0,1$, see
Paper I) the expansion in $\zeta$ is accurate within $1\%$ as long as $\zeta<0.4$ for the real parts of the modes, and
$\zeta<0.3$ for the imaginary parts. Thus, in the rest of the paper we shall consider these ranges for the coupling $\zeta$.
In Fig.~\ref{fig:truncerr-w2} we also show the relative shift between the functions $\omega^{nl}_{r}$ in GR and in EdGB
gravity; we can see that the truncation error $\epsilon^{nl(s)}_r$ at $s=5$ is significantly smaller than the EdGB
contribution.

\begin{figure*}[hpt]
	\includegraphics[width=8cm]{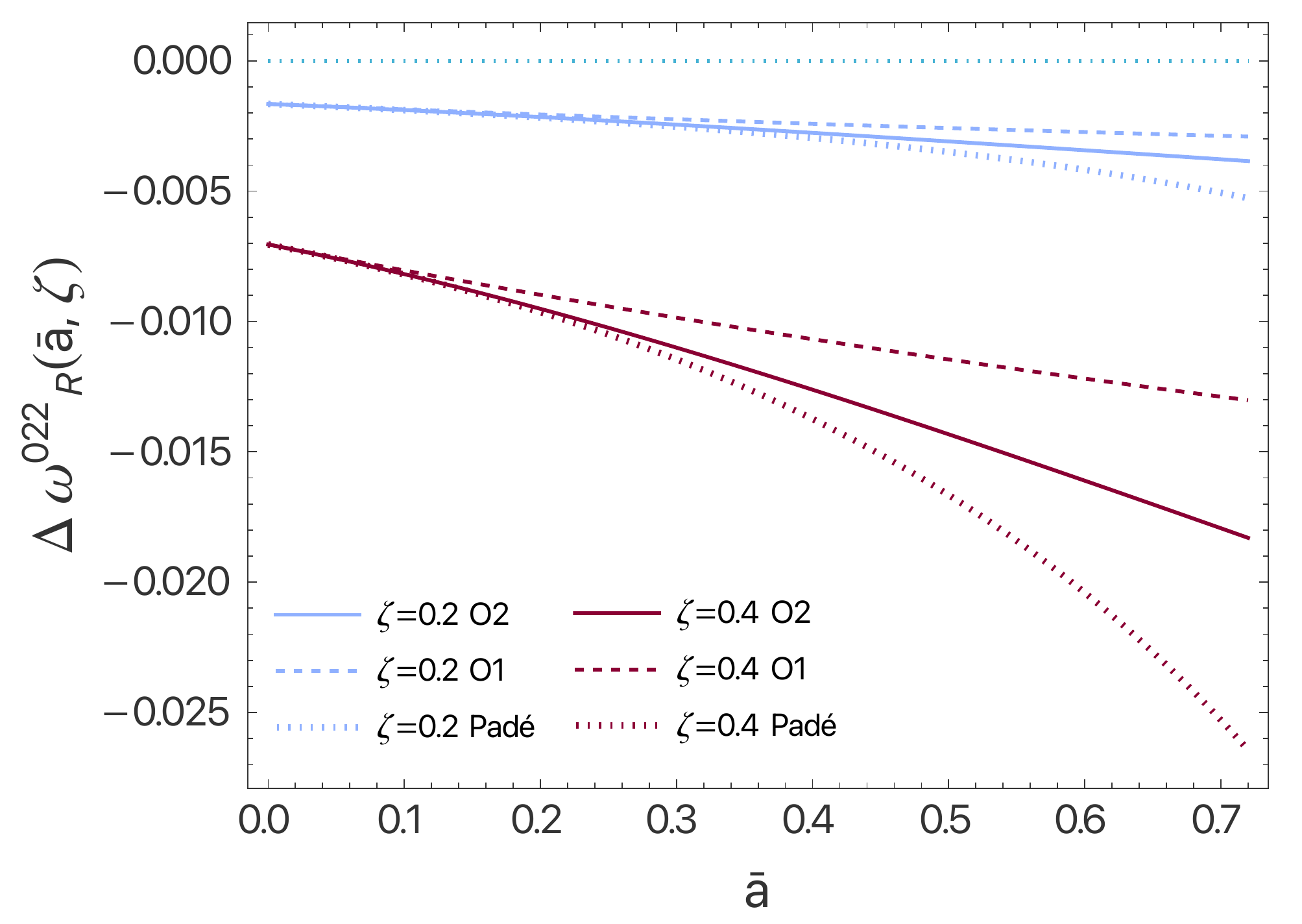}
	\includegraphics[width=8cm]{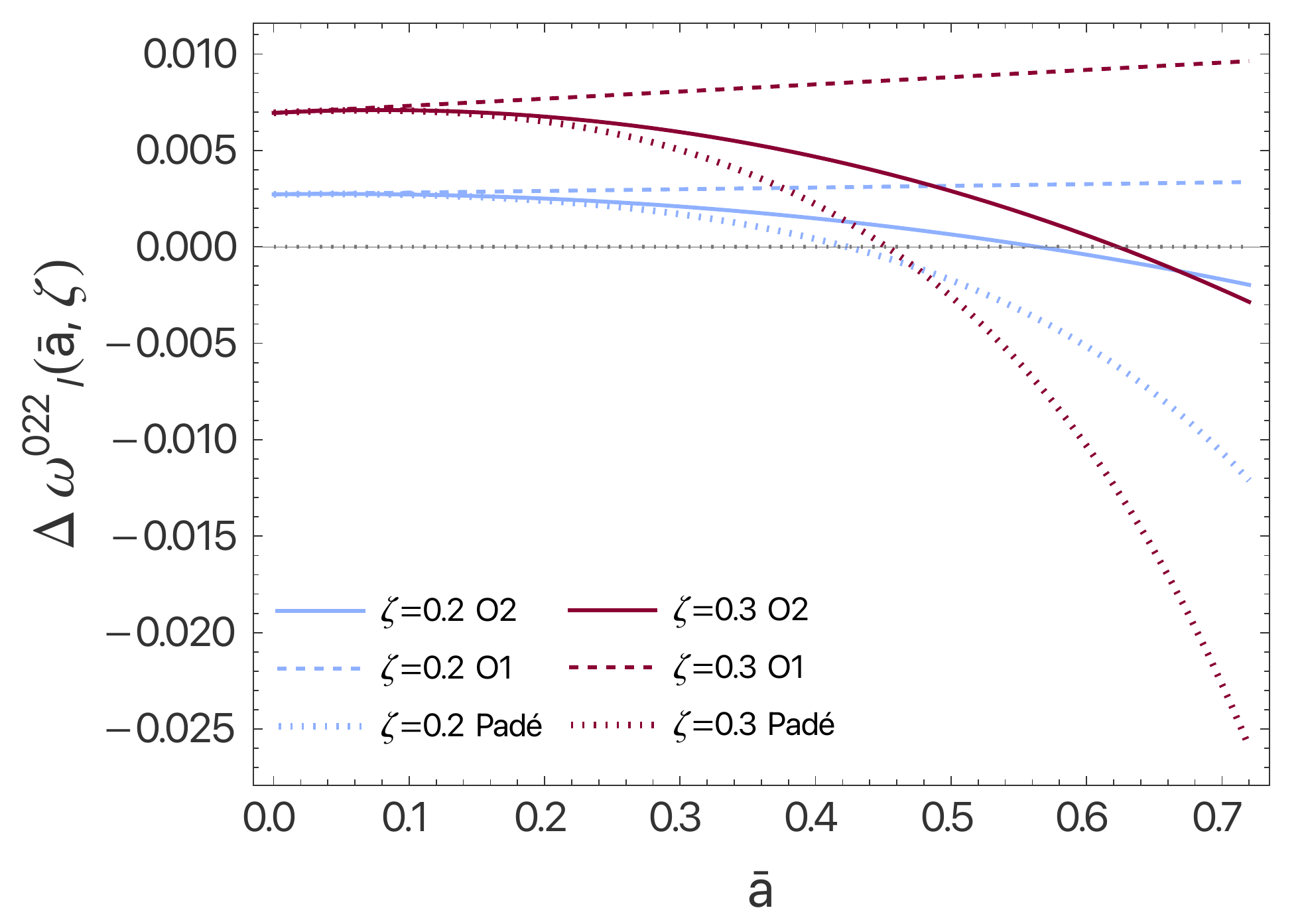}
	\includegraphics[width=8cm]{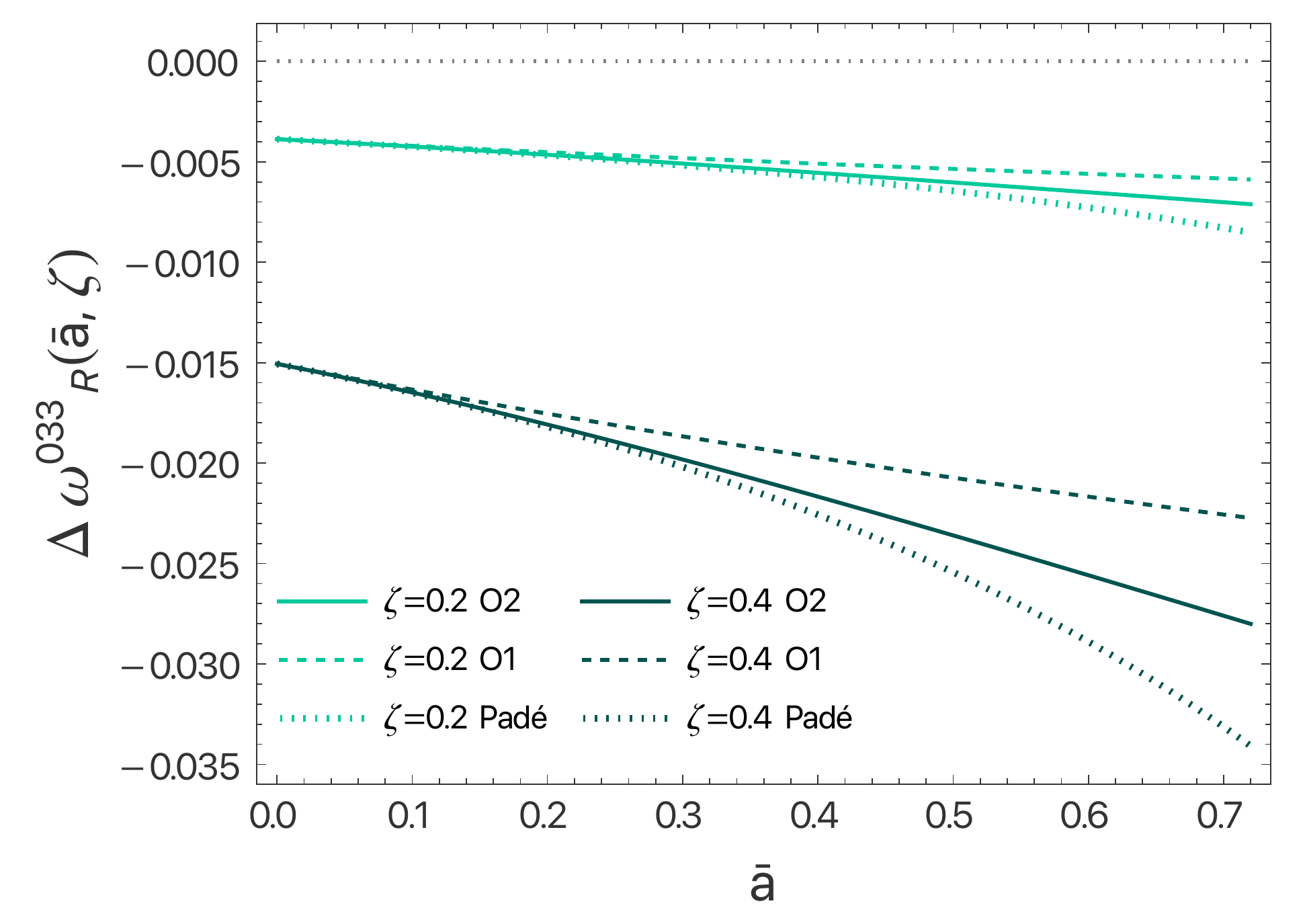}
	\includegraphics[width=8cm]{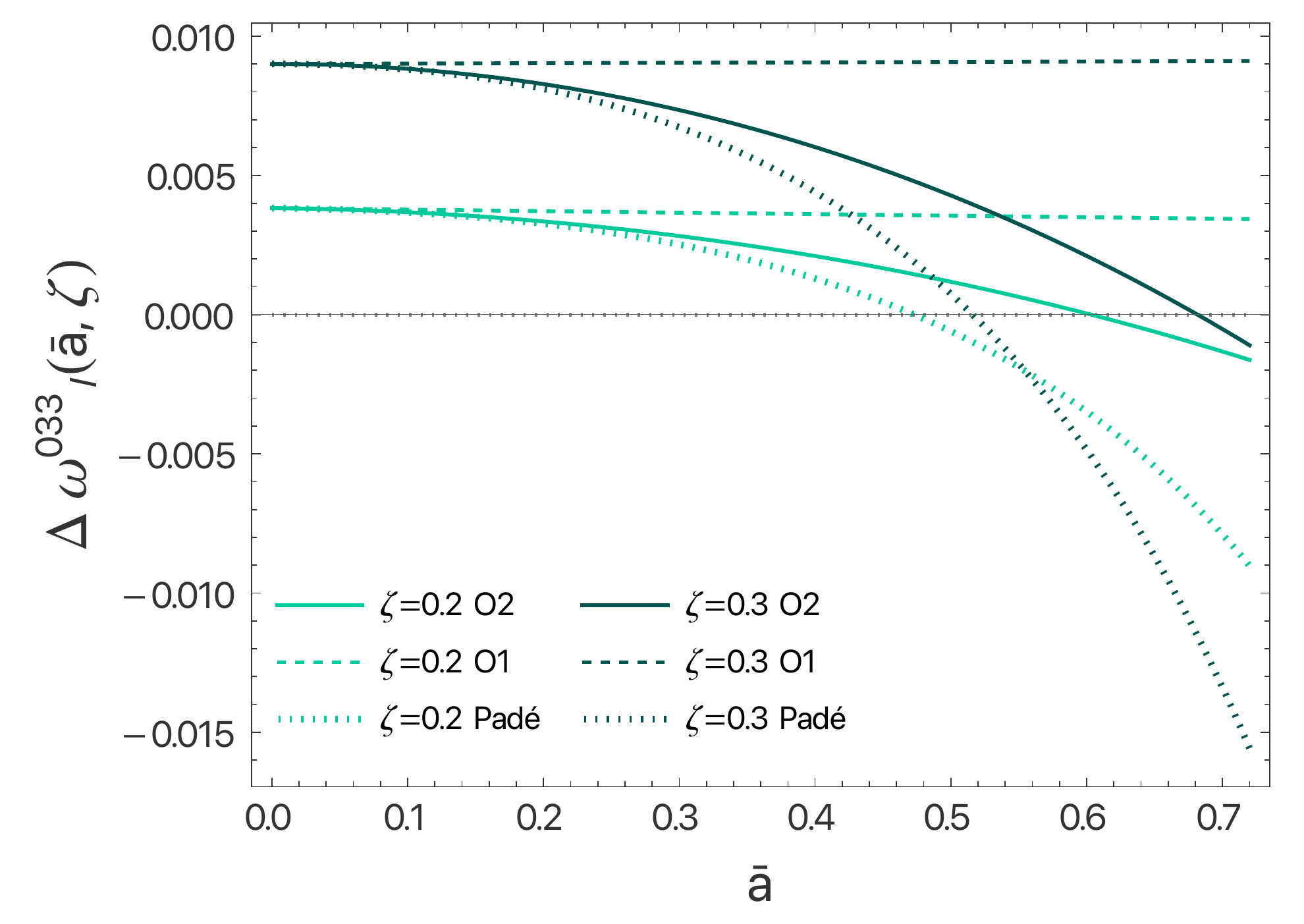}
	\caption{Real (left panels) and imaginary (right panels) parts of the relative difference of EdGB QNMs with
          respect to GR, as a function of $\bar a$. We consider the $(nlm)=(022)$ (upper panels) and $(nlm)=(033)$
          (lower panels) QNMs computed up to the first order in the spin, up to second order, and with Pad\'e
          resummation.}
	\label{fig:delta-wtot}
\end{figure*}
\begin{figure*}[hpt]
	\includegraphics[width=8cm]{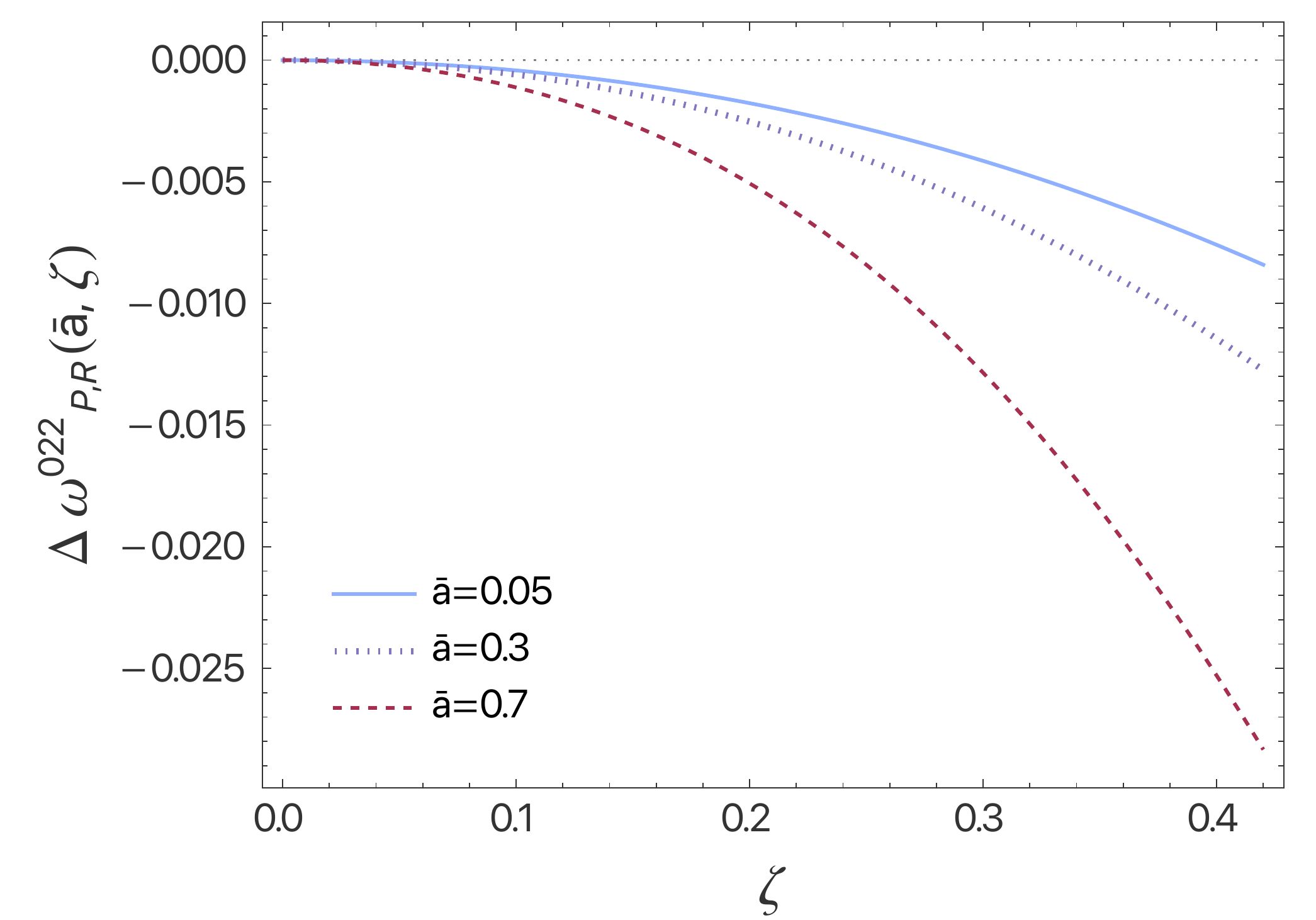}
	\includegraphics[width=8cm]{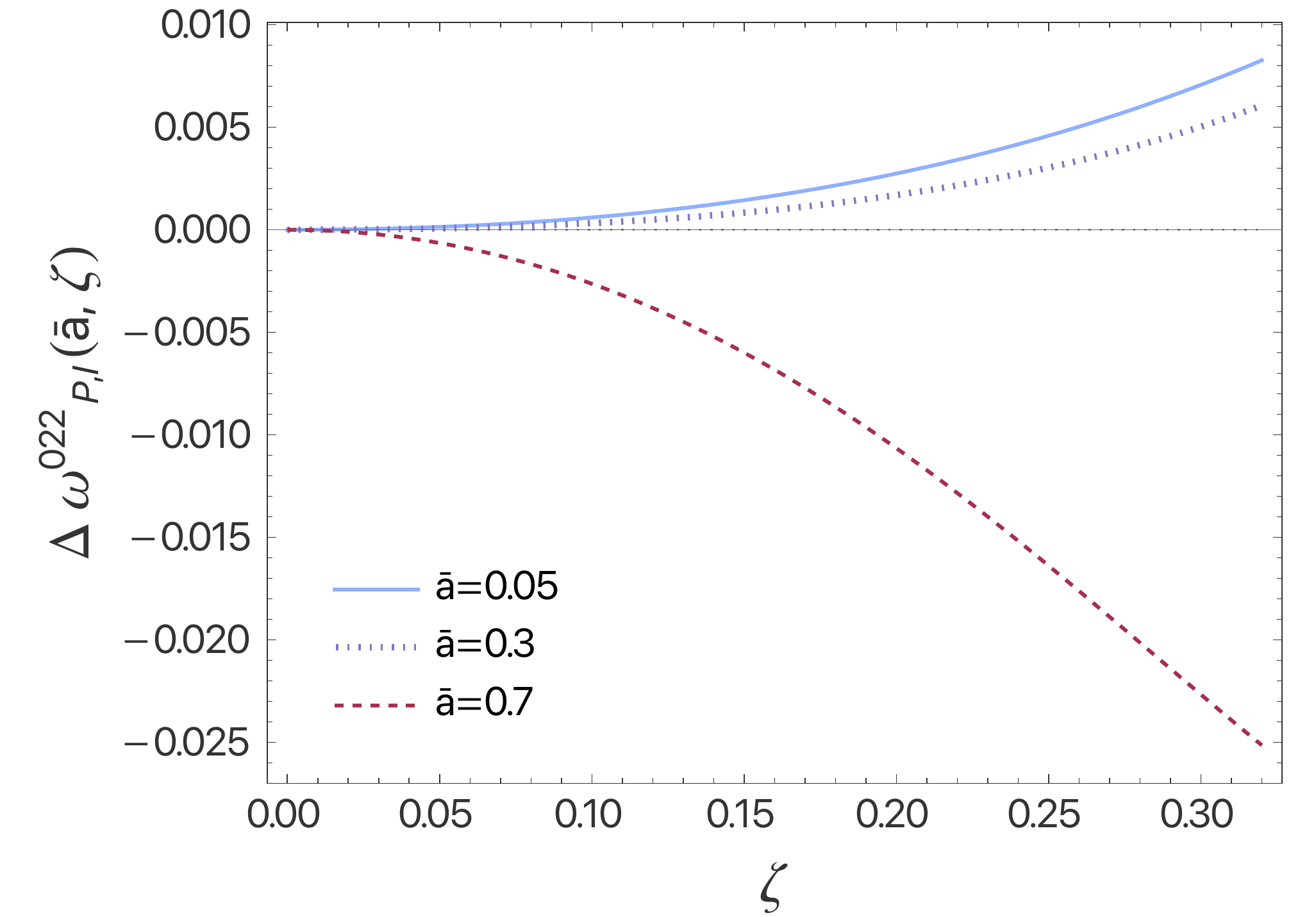}
	\includegraphics[width=8cm]{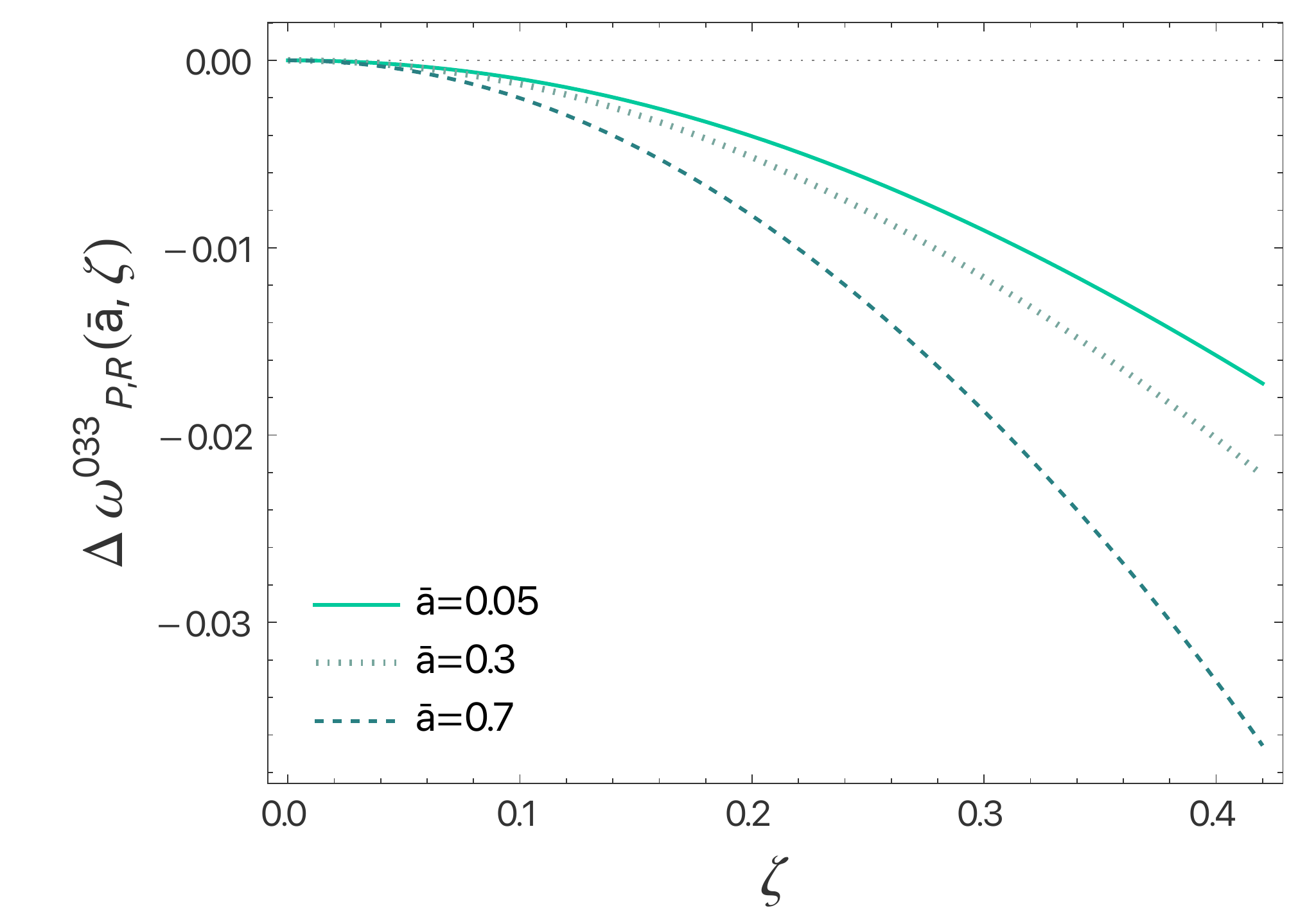}
	\includegraphics[width=8cm]{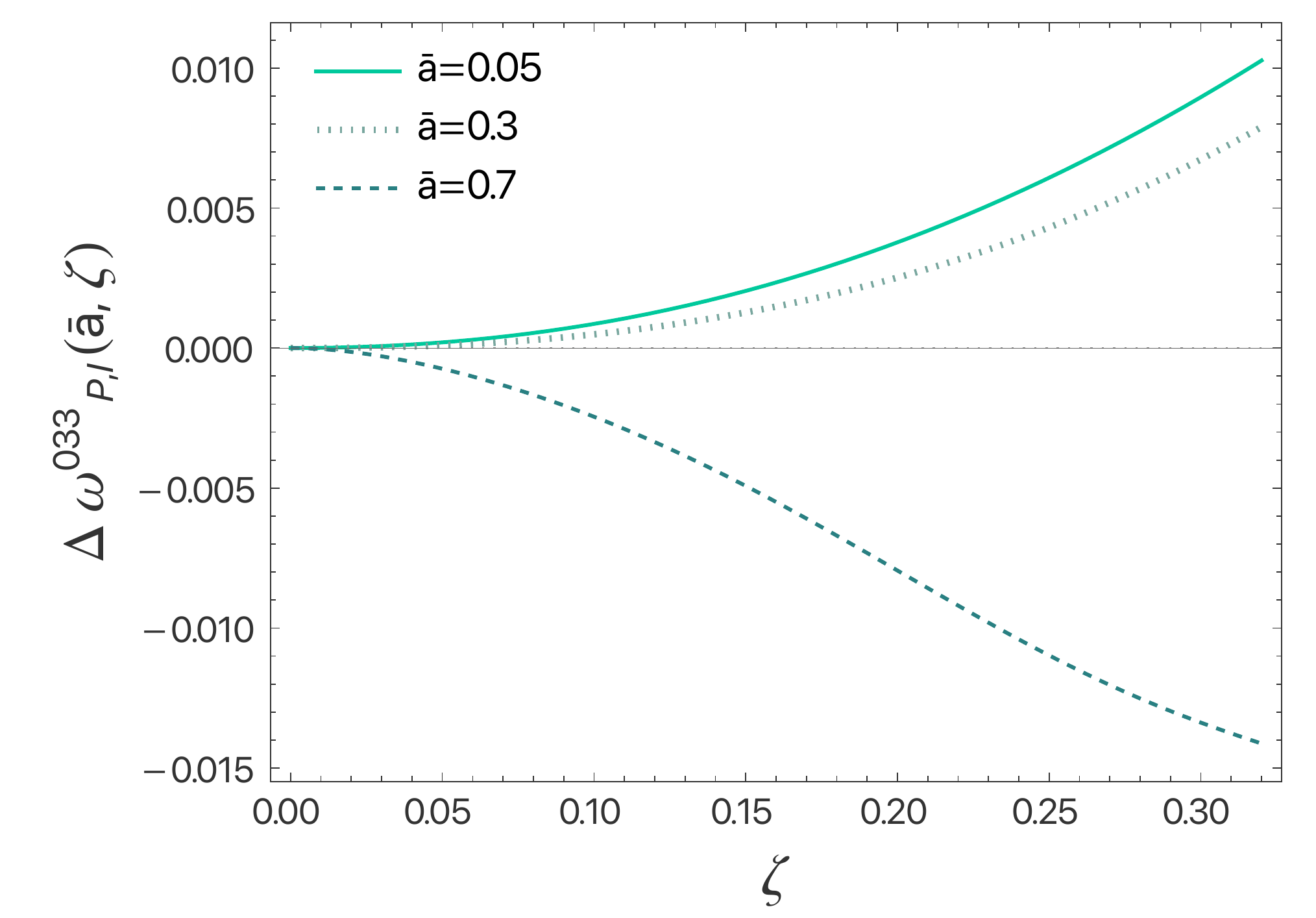}
	\caption{Same as Fig.~\ref{fig:delta-wtot}, as a function of $\zeta$, with QNMs computed with Pad\'e
          resummation.}
	\label{fig:delta-wtot-a-fixed}
\end{figure*} 
%

\subsection{Quasi-normal modes}
%
The Pad\'e QNMs with $(nlm)=(022),\,(033)$ are shown in Fig.~\ref{fig:wtot}, as functions of the spin, for different
values of $\zeta$.  We can see
that (for these modes) while at low values of the spin the EdGB corrections increase (in modulus) both the real and
imaginary parts of the QNMs, when the spin is larger the EdGB correction increases the real parts of the modes,
decreases the imaginary parts.

In order to undestand the effect of the EdGB corrections, it is useful to define the relative differences between the
QNMs in EdGB gravity and in GR:
\begin{equation}
  \Delta \omega^{nlm}_{R,I}(\bar a,\zeta) = \frac{\omega^{nlm}_{R,I}(\bar a,\zeta)
    -\omega^{nlm}_{R,I}(\bar a,0)}{\omega^{nlm}_{R,I}(\bar a,0)}
\end{equation}
where $R,I$ refer to the real and imaginary parts, respectively. These quantities are shown in
Fig.~\ref{fig:delta-wtot}, for different values of $\zeta$, as functions of $\bar a$, and in
Fig.~\ref{fig:delta-wtot-a-fixed} for different values of $\bar a$, as functions of $\zeta$. The spin expansion is
performed to first and second order, and resummed using Pad\'e approximants; in Fig.~\ref{fig:delta-wtot-a-fixed} it is
only resummed using Pad\'e approximants.

We note that (as argued in Paper I) the $O({\bar a^2})$ terms enhance the EdGB corrections to the QNMs; moreover, the
corrections are further enhanced by the Pad\'e resummation. For $\bar a=0.7$, the $l=2$ fundamental mode is shifted of
$\sim0.5\%$ for $\zeta=0.2$, and of $\sim2.5\%$ for $\zeta=0.4$.
We also note that the EdGB relative corrections of the imaginary parts change sign for large values of the spins; this
explain the decreasing of the EdGB correction discussed above.
\begin{figure*}[hpt]
	\includegraphics[width=8cm]{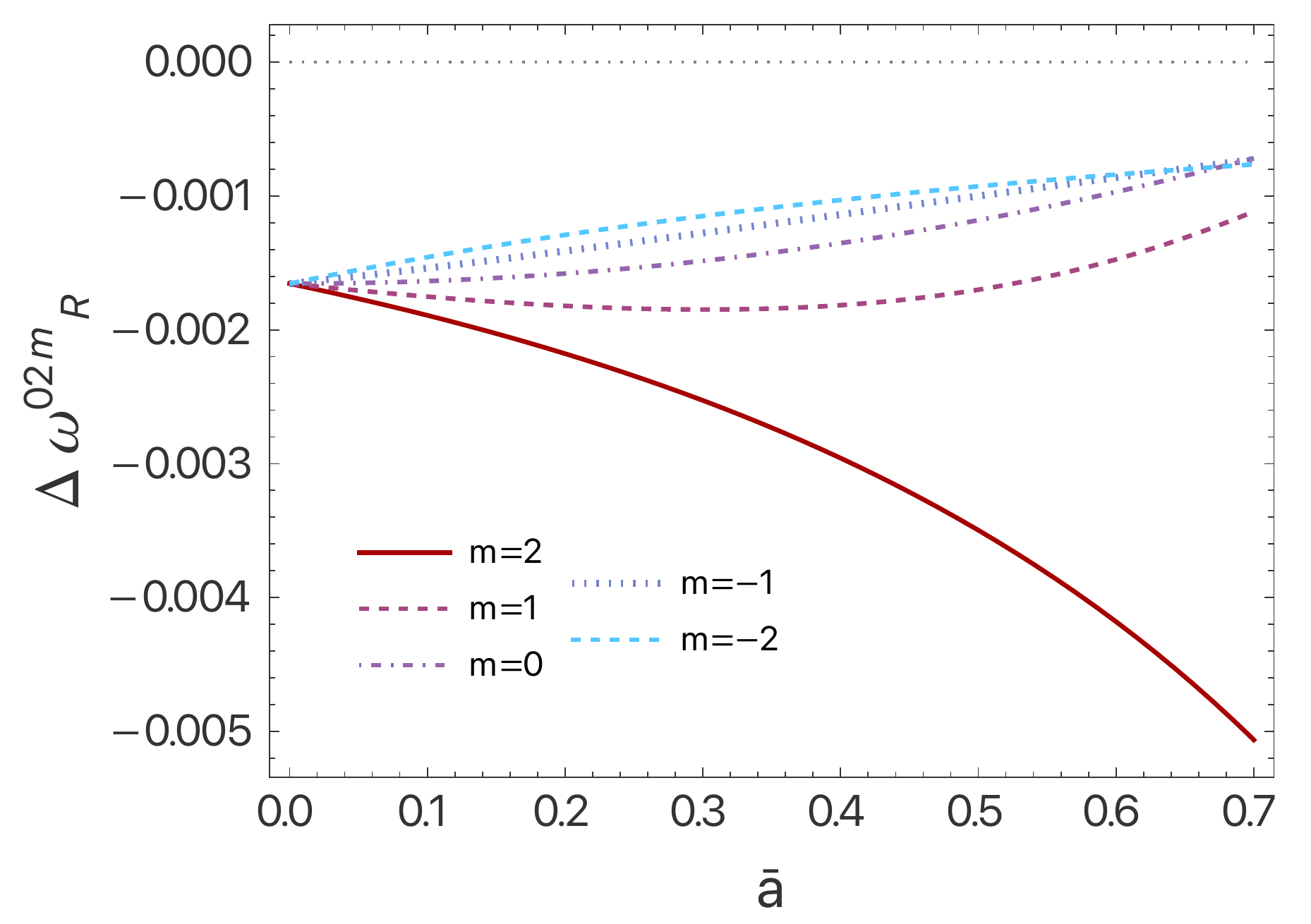}
	\includegraphics[width=8cm]{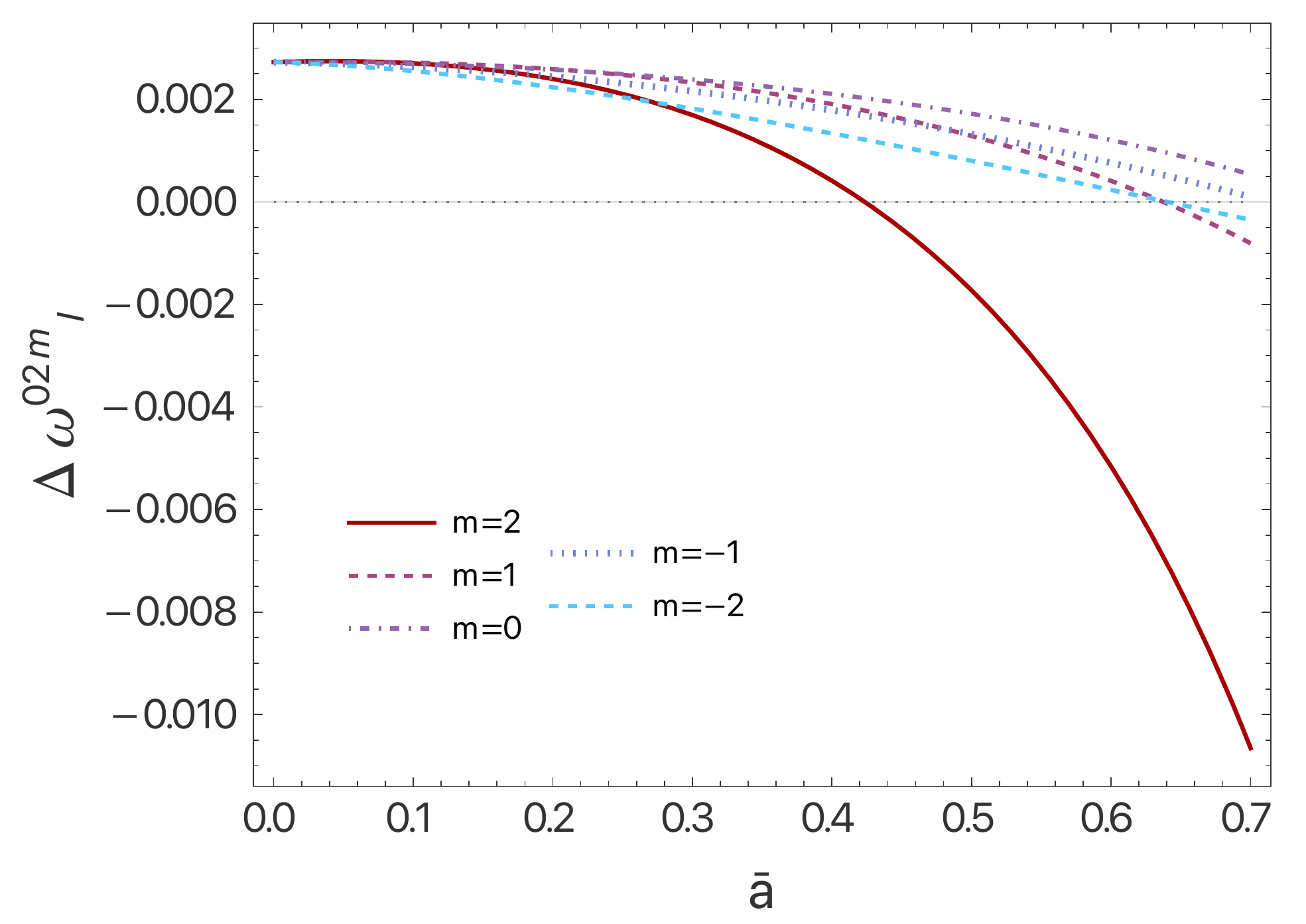}
	\includegraphics[width=8cm]{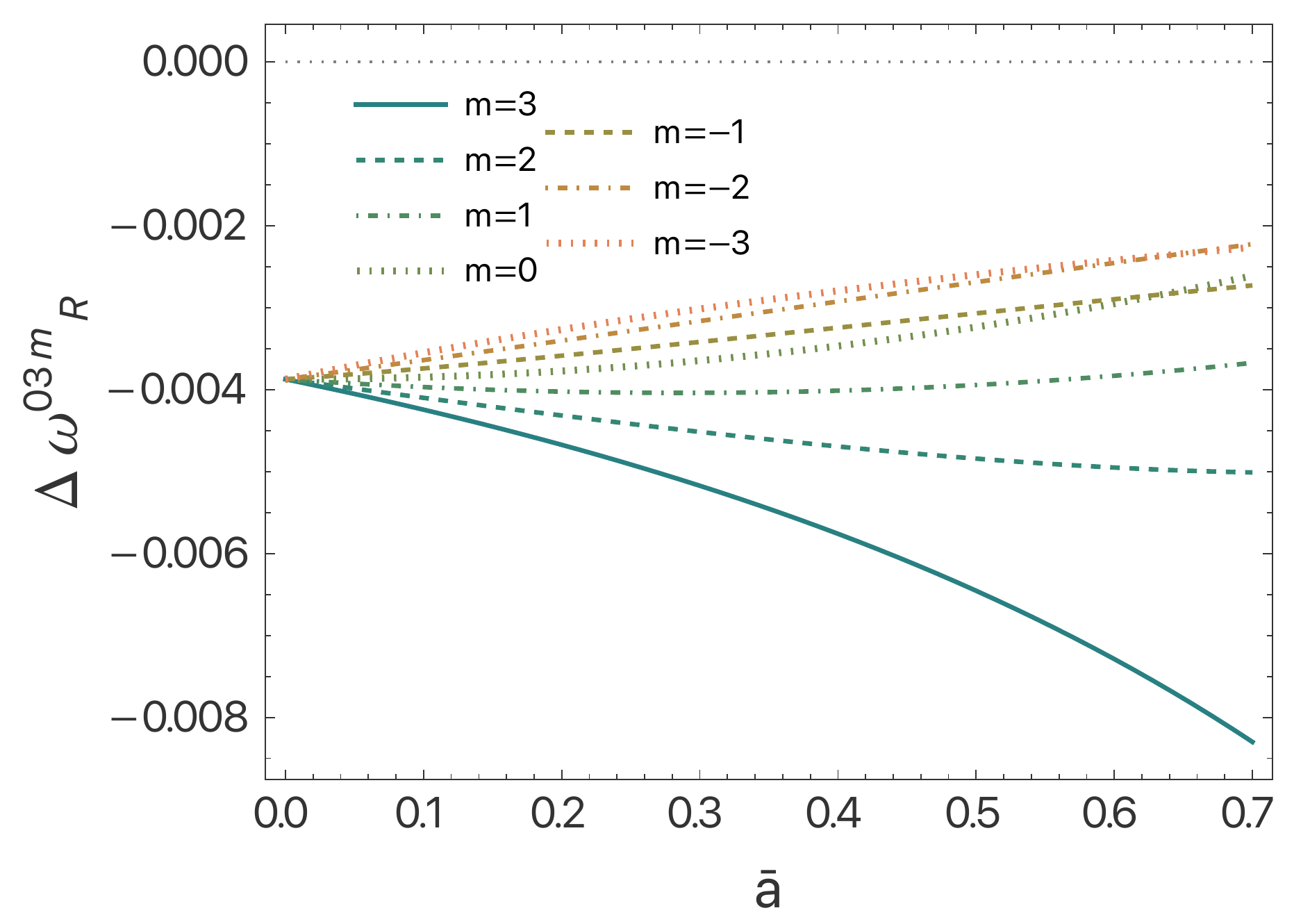}
	\includegraphics[width=8cm]{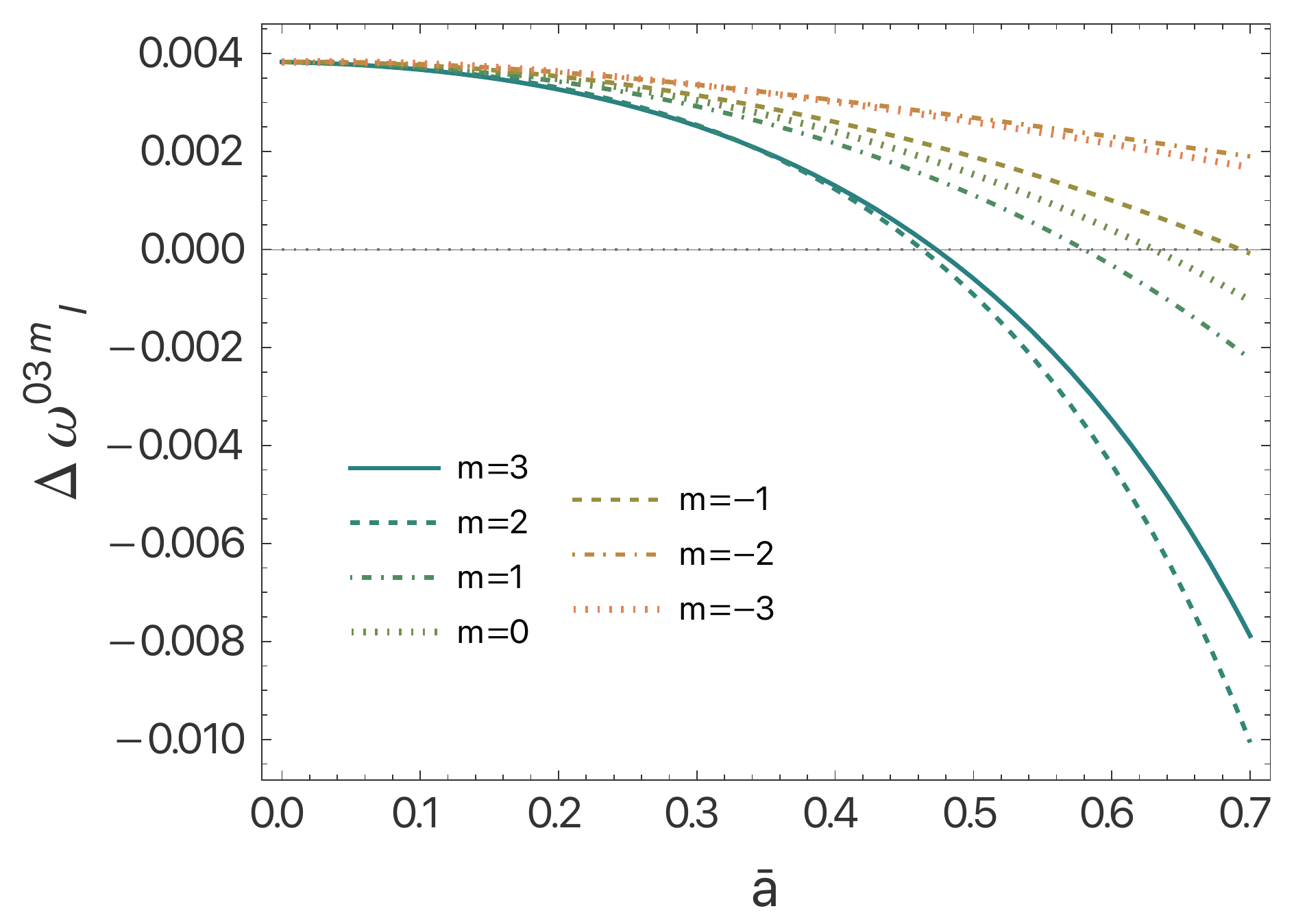}
	\caption{Real (upper panels) and imaginary (lower panels) of the relative difference of EdGB QNMs with respect
          to GR as a function of $\bar a$, for $\bar\zeta=0.2$ and different values of $m$. We consider $(nl)=(02)$
          (left panels) and $(nl)=(03)$ (right panels) QNMs.}
	\label{fig:delta-wtot-diversim}
\end{figure*}

From Fig.~\ref{fig:delta-wtot-a-fixed} we note that when $\zeta\lesssim0.3$, the contribution of the GR deviations to
the QNMs is typically smaller than $2\%$, which is the error we expect from the slow-rotation expansion (see
Fig.~\ref{fig:delta-kerr-SR}). Therefore, the slow-rotation expansion to $O({\bar a}^2)$ discussed in this paper should
 only be used for the EdGB corrections ($O(\zeta^2)$), while Kerr modes (or a slow-rotation expansion to a high order)
should be used to compute the GR contribution ($O(\zeta^0)$). In this way, the error due to the $O({\bar a}^2)$
truncation would only affect the EdGB part of the modes.
Finally, in Fig.~\ref{fig:delta-wtot-diversim} we show the EdGB relative corrections for the fundamental modes with
$l=2,3$, for different values of $m$.
%
\subsection{Fits and Taylor expansions in the coupling constant}
%
We have fitted the functions $\omega^{nl}_{r}(\zeta)$ defined in Eq.~\eqref{eq:qnms-expansion-2} with sixth-order
polynomials in $\zeta$ ($\zeta\in[0.0.4]$ for the real parts, $\zeta\in[0,0.3]$ for the imaginary parts):
\begin{align}
M \, \omega^{nl}_{r} (\zeta) = \sum_{i=0}^{6} \zeta^i C^{nl}_{r\,i}\,.
\label{eq:eq-coeff-general}
\end{align}
Since for gravitational-led modes the EdGB correction is of $O(\zeta^2)$, we have set
$C^{nl}_{r\,1}=0$\,\cite{Salcedo2016}.
We have estimated the relative error of the fit\,\eqref{eq:eq-coeff-general}, $\delta_{\rm f}$, as the mean over 100
attempts of the relative difference between the fit, computed from randomly selected $80 \%$ of the data points, and the
remaining $20\%$ of the data. For $l=3$ and $r=2a\,2b$, the functions have been fitted with fourth-order polynomials,
because the error $\delta_{\rm f}$ is smaller.

In Tables \ref{tab:coeffs-fit-w0}-\ref{tab:coeffs-fit-w2b} we show the coefficients of the
fit~\eqref{eq:eq-coeff-general} for the (gravitational-led, polar-led) fundamental modes with $l=2,3$. We also show the
corresponding relative errors $\delta_{\rm f}$.
\begin{table}[htp]
  \caption{\textit{Coefficients of the fit~\ref{eq:eq-coeff-general} of $\omega^{0l}_{r}$ for gravitational-led,
      polar-led modes, with $r=0$, $l=2,3$, up to $i=6$. In the last line we show the relative error of the fit,
      $\delta_{\rm f}$.}} 
    \centering
	\begin{tabular}{ c | c c | c c }
		\toprule
		 & Re ($ {l} =2$) & Im ($ {l} =2$) & Re ($ {l} =3$) & Im ($ {l} =3$)  \\
		\midrule
		$C_0$ & 0.37367 & $-0.08896$ & 0.59944 & $-0.09270$  \\
		$C_2$ & $-1.406 \cdot 10^{-2}$ & $-4.70 \cdot 10^{-3}$ & $-5.453 \cdot 10^{-2}$ & $-7.19 \cdot 10^{-3}$  \\
		$C_3$ &  $-7.53 \cdot 10^{-3}$ & $-6.10 \cdot 10^{-3}$ & $-3.093 \cdot  10^{-2}$ & $-1.098 \cdot 10^{-2}$  \\
		$C_4$ &  $1.35 \cdot 10^{-3}$ & $-3.22 \cdot 10^{-3}$ & $6.419 \cdot 10^{-2}$ & $1.001 \cdot 10^{-2}$ \\
		$C_5$ &  $7.09 \cdot 10^{-3}$ & $-1.61 \cdot 10^{-3}$ & $2.417 \cdot 10^{-2}$ & $2.158 \cdot 10^{-2}$ \\
		$C_6$ &  $-2.03 \cdot 10^{-3}$ & $-2.8 \cdot 10^{-4}$ & $-5.215 \cdot 10^{-2}$ & $-3.345\cdot 10^{-2}$ \\
		\midrule
		$\delta_{\rm f}$ &  $2 \cdot 10^{-9}$ & $3 \cdot 10^{-8}$ & $10^{-7}$ & $2 \cdot 10^{-7}$ \\
		\bottomrule
	\end{tabular}
	\label{tab:coeffs-fit-w0}
\end{table}
\begin{table}[htp]
	\caption{\textit{As in Table\,\ref{tab:coeffs-fit-w0}, with $r=1$.}}
	\centering
	\begin{tabular}{ c | c c | c c }
		\toprule
		& Re ($ {l} =2$) & Im ($ {l} =2$) & Re ($ {l} =3$) & Im ($ {l} =3$)  \\
		\midrule
		$C_0$ & 0.06289 & $0.00100$ & 0.06737 & $0.00065$  \\
		$C_2$ & $-1.048 \cdot 10^{-2}$ & $2 \cdot 10^{-5}$ & $-2.156 \cdot 10^{-2}$ & $1.18 \cdot 10^{-3}$  \\
		$C_3$ &  $-1.074 \cdot 10^{-2}$ & $-2.69 \cdot 10^{-3}$ & $-2.056 \cdot  10^{-2}$ & $-3.19 \cdot 10^{-3}$  \\
		$C_4$ &  $-1.53 \cdot 10^{-3}$ & $-9.86 \cdot 10^{-3}$ & $4.465 \cdot 10^{-2}$ & $3.9 \cdot 10^{-4}$ \\
		$C_5$ &  $-2.40 \cdot 10^{-3}$ & $8.90 \cdot 10^{-3}$ & $2.341 \cdot 10^{-2}$ & $-1.224 \cdot 10^{-2}$ \\
		$C_6$ &  $1.433 \cdot 10^{-2}$ & $-3.773 \cdot 10^{-2}$ & $-6.800 \cdot 10^{-2}$ & $1.056 \cdot 10^{-2}$ \\
		\midrule
		$\delta_{\rm f}$ &  $5 \cdot 10^{-8}$ & $2 \cdot 10^{-5}$ & $10^{-6}$ & $10^{-4}$ \\
		\bottomrule
	\end{tabular}
	\label{tab:coeffs-fit-w1}
\end{table}
\begin{table}[htp]
	\caption{\textit{As in Table\, \ref{tab:coeffs-fit-w0}, with $r=2a$. The fit for
            $l=3$ stops at fourth order in $\zeta$.}}
	\centering
	\begin{tabular}{ c | c c | c c }
		\toprule
		 & Re ($ {l} =2$) & Im ($ {l} =2$) & Re ($ {l} =3$) & Im ($ {l} =3$)  \\
		\midrule
		$C_0$ & 0.03591 & $0.00638$ & 0.04755 & $0.00659$  \\
		$C_2$ & $1.348 \cdot 10^{-2}$ & $6.50 \cdot 10^{-3}$ & $2.941 \cdot 10^{-2}$ & $1.857 \cdot 10^{-2}$  \\
		$C_3$ &  $1.051 \cdot 10^{-2}$ & $8.48 \cdot 10^{-3}$ & $2.354 \cdot  10^{-2}$ & $8.93 \cdot 10^{-3}$  \\
		$C_4$ &  $1.051 \cdot 10^{-2}$ & $5.06 \cdot 10^{-3}$ & $-2.391 \cdot 10^{-2}$ & $-2.78 \cdot 10^{-3}$ \\
		$C_5$ &  $4.85 \cdot 10^{-3}$ & $1.402 \cdot 10^{-2}$ & $-$ & $-$ \\
		$C_6$ &  $1.037 \cdot 10^{-2}$ & $-2.24 \cdot 10^{-3}$ & $-$ & $-$ \\
		\midrule
		$\delta_{\rm f}$ &  $7 \cdot 10^{-7}$ & $2 \cdot 10^{-6}$ & $10^{-4}$ & $10^{-3}$ \\
		\bottomrule
	\end{tabular}
	\label{tab:coeffs-fit-w2a}
\end{table}
\begin{table}[htp]
	\caption{\textit{As in Table\,\ref{tab:coeffs-fit-w2a}, with $r=2b$.}}
	\centering
	\begin{tabular}{ c | c c | c c }
		\toprule
		& Re ($ {l} =2$) & Im ($ {l} =2$) & Re ($ {l} =3$) & Im ($ {l} =3$)  \\
		\midrule
		$C_0$ & 0.00896 & $-0.00031$ & 0.00661 & $0.00006$  \\
		$C_2$ & $-8.37 \cdot 10^{-3}$ & $3.13 \cdot 10^{-3}$ & $-9.95 \cdot 10^{-3}$ & $7.0 \cdot 10^{-4}$  \\
		$C_3$ &  $-1.201 \cdot 10^{-2}$ & $2.95 \cdot 10^{-3}$ & $-4.90 \cdot  10^{-3}$ & $-2.27 \cdot 10^{-3}$  \\
		$C_4$ &  $2.67 \cdot 10^{-3}$ & $-1.046 \cdot 10^{-2}$ & $4.78 \cdot 10^{-3}$ & $-1.57 \cdot 10^{-3}$ \\
		$C_5$ &  $-5.926 \cdot 10^{-2}$ & $3.088 \cdot 10^{-2}$ & $-$ & $-$ \\
		$C_6$ &  $8.254 \cdot 10^{-2}$ & $-6.819 \cdot 10^{-2}$ & $-$ & $-$ \\
		\midrule
		$\delta_{\rm f}$ &  $5 \cdot 10^{-6}$ & $5 \cdot 10^{-4}$ & $2 \cdot 10^{-4}$ & $4 \cdot 10^{-2}$ \\
		\bottomrule
	\end{tabular}
	\label{tab:coeffs-fit-w2b}
\end{table}
\begin{table}
  \caption{\textit{Coefficients $\omega^{nl}_{r\,(A)}$  of the Taylor expansion~\eqref{eq:taylorzeta},
      for the gravitational-led, polar-led fundamental ($n=0$) modes with $l=2,3$.}}  \centering
	\begin{tabular}{ c | c c | c c }
		\toprule 
		$r\,A$ & Re ($ {l} =2$) & Im ($ {l} =2$) & Re ($ {l} =3$) & Im ($ {l} =3$)  \\
		\midrule
		$0\,0$ & 0.37367 & $-0.08896$ & 0.59944 & $-0.09270$  \\
		$1\,0$ & $6.289\cdot 10^{-2}$ & $1.00 \cdot 10^{-3}$ & $6.737 \cdot 10^{-2}$
                & $6.5 \cdot 10^{-4}$  \\
		$2a\,0$ &  $3.591 \cdot 10^{-2}$ & $6.38 \cdot 10^{-3}$ & $4.755\cdot  10^{-2}$ &
                $6.59 \cdot 10^{-3}$  \\
		$2b\,0$ &  $8.96 \cdot 10^{-3}$ & $-3.1 \cdot 10^{-4}$ & $6.61 \cdot 10^{-3}$
                & $6 \cdot 10^{-5}$ \\
		\bottomrule
                \toprule
		$r\,A$ & Re ($ {l} =2$) & Im ($ {l} =2$)  & Re ($ {l} =3$) & Im ($ {l} =3$)    \\		
		\midrule
		$0\,2$ & $-1.411 \cdot 10^{-2}$ & $-4.70 \cdot 10^{-3}$ & $-5.463 \cdot 10^{-2} $
                & $-7.21 \cdot 10^{-3}$  \\
		$1\,2$ & $-1.049 \cdot 10^{-2}$ & $4 \cdot 10^{-5}$ & $-2.166 \cdot 10^{-2}$
                & $1.17 \cdot 10^{-3}$  \\
		$2a\,2$ &  $1.340 \cdot 10^{-2}$ & $6.54 \cdot 10^{-3}$ & $2.947 \cdot 10^{-2}$
                & $1.777 \cdot 10^{-2}$ \\
		$2b\,2$ &  $-8.42 \cdot 10^{-3}$ & $3.19 \cdot 10^{-3}$  & $-9.50 \cdot 10^{-3}$
                & $4.3 \cdot 10^{-4}$\\
		\bottomrule
	\end{tabular}
	\label{tab:SRSC-expa-coeff}
\end{table}

These fits are very accurate to describe the functions $\omega^{nl}_{r} (\zeta)$ in the entire range $\zeta\in[0,0.4]$
($[0,0.3]$ for the imaginary parts). If we are interested in these functions for $\zeta\ll1$, we should instead compute
a {\it Taylor expansion} of them around $\zeta=0$. A Taylor expansion is also useful for data analysis techniques based
on QNM expansions in the spin, like {\sc ParSpec }\,\cite{Maselli:2019mjd,Carullo:2021dui,parspec2}. Therefore, we
performed a Taylor expansion of  the functions $\omega^{nl}_{r} (\zeta)$ to $O(\zeta^2)$:
\begin{equation}\label{eq:taylorzeta}
  \omega^{nl}_r(\zeta)=  \omega^{nl}_{r\,(0)}+\zeta^2 \omega^{nl}_{r\,(2)}+O(\zeta^3)
  \end{equation}
(as mentioned above, since we are considering gravitational-led modes, the first-order contributions $
\omega^{nl}_{r\,(1)}$ identically vanish).
The coefficients of the expansion~\eqref{eq:taylorzeta}, for the $(nl)=(02),\,(03)$ modes, are given in
Table~\ref{tab:SRSC-expa-coeff}.

\section{Conclusions and outlook}\label{sec:concl}
In this article we have computed the QNMs of a rotating BH in EdGB gravity. Strictly speaking, this is a slow-rotation
computation, since it is based on an expansion in the spin $\bar a$ up to second order. However, the use of Pad\'e
approximants enhances the range of validity of the expansion: an analysis of the general relativistic case suggests that
the QNMs derived with this approach are accurate within $\sim2\%$ up to phenomenologically relevant values of the spin,
i.e. for $\bar a\lesssim0.7$ (see also\,\cite{Hatsuda:2020egs}).

We find that (as argued in Paper I) the second-order contribution greatly enhances the EdGB correction to the QNMs. For
instance, for the real part of the $(nlm)=(022)$ -- which is the mode typically excited with largest amplitude in actual
BH ringdowns (see e.g.~\cite{Ghosh:2021mrv}) -- , assuming a BH spin of $\bar a=0.7$, the EdGB correction estimated to
$O(\bar a)$ for $\zeta\simeq 0.4$ ($\zeta\simeq0.3$ for the imaginary part) is of $\simeq1\%$, while that estimated to
$O({\bar a}^2)$ (with Pad\'e resummation) is of $\simeq2.5\%$ (see Fig.~\ref{fig:delta-wtot-a-fixed}).\footnote{Note that
  in the Conclusions of Paper I, due to a typographical error, we wrote that the EdGB correction of the $(022)$ mode for
  a BH with $\bar a=0.7$, estimated to first order in the spin, is $18\%$ while the correct number was $1.8\%$.}

Our computation has been performed expanding the background and the perturbation equations in the coupling constant to
$O(\zeta^6)$. An analysis of the truncation error indicates that our results are accurate for $\zeta\lesssim0.4$ for the
real parts of the modes, $\zeta\lesssim0.3$ for the imaginary parts. We provide analytical fits of the modes in this
range of the coupling constant. We also provide a Taylor expansion around $\zeta=0$, which can be useful in the data
analysis of ringdown signals.

Concerning the detectability of the EdGB deviations, we note that $O(10)$ detections of binary BH ringdowns with
signal-to-noise ratios (SNRs) of the order of $\sim30$ are expected to be sufficient to measure BH QNMs with an accuracy
of few percent (see e.g.~\cite{Brito:2018rfr}). Since third-generation detectors, like the Einstein
Telescope\,\cite{Punturo:2010zz}, are expected to reach even larger SNRs, they could be sensitive enough to find the
deviations studied in this article, at least for the largest values of the coupling. This, however, is just an
order-of-magnitude estimate: in order to assess the detectabilty of EdGB corrections in the BH QNMs by third-generation detectors
we need to know the SNR and the number of events required to measure the EdGB shifts, as functions of the coupling
constant. This can only be found with a proper sensitivity analysis of the combined detections of several oscillating
BHs, with different masses and spins. Such analysis, based on the {{\sc ParSpec }} framework\,\cite{Maselli:2019mjd}, is
currently in preparation\,\citep{parspec2}.

This is the first computation, in a modified gravity theory, of the QNMs of BHs to second order in rotation.  Although
EdGB gravity is an interesting theory by itself for a number of reasons, this can also be considered as a study case, to
understand which kind of deviation we may expect in the ringdown signal. The mode corrections in specific theories of
gravity are a necessary ingredient of gravitational spectroscopy, using future GW data to perform tests of gravity which
go beyond null tests of GR\,\cite{Maselli:2019mjd,Carullo:2021dui,parspec2}.  Of course, the next step will to extend
this computation to other classes of possible GR deviations.

The computation presented here will also be useful, once fully numerical simulations of BH coalescences in EdGB gravity will be
available\,\cite{Witek:2018dmd,Okounkova:2020rqw,Witek:2020uzz,East:2020hgw}, as a benchmark to test the numerical
codes.

\begin{acknowledgments}
  We are indebted to Emanuele Berti for suggesting us the use of Pad\'e approximants. We also thank Paolo Pani, Andrea
  Maselli and Hector Silva for useful discussions. We acknowledge networking support by the COST Action CA16104. We also
  acknowledge support from the Amaldi Research Center funded by the MIUR programs "Dipartimento di Eccellenza" (CUP:
  B81I18001170001) and PRIN2017-MB8AEZ.  We acknowledge financial support from the EU Horizon 2020 Research and
  Innovation Programme under the Marie Sklodowska-Curie Grant Agreement no.~101007855.
\end{acknowledgments}

\clearpage
\appendix 

\section{Equations for gravitational perturbations at second order in the spin}\label{app:eq}
The field equations~\eqref{eq:scalar}, \eqref{eq:metric}, linearized in the perturbation around the stationary BH
solution discussed in Sec.~\ref{sec:stationaryebhdgb}, can be written as follows (we follow the same notation
as~\cite{Kojima1992}, and leave implicit the sum over $l,m$):
\begin{widetext}
\begin{align}
  \left[A^{(I)}_{lm}+ \tilde{A}^{(I)}_{lm} \cos\theta + \hat{A}^{(I)}_{lm} \cos^2\theta \right]Y(\theta) +  i m \left[ C^{(I)}_{lm}
    + \tilde{C}^{(I)}_{lm} \cos\theta \right] Y^{lm}(\theta) + m^2 E^{(I)}_{lm} Y^{lm}(\theta)&   \notag \\
     +  \left[B^{(I)}_{lm}+\tilde{B}^{(I)}_{lm} \cos\theta + i m D^{(I)}_{lm} \right]\
  \sin\theta \ {Y'}^{lm}(\theta)&=0\,, 
\label{eq:einst-first-group}\\
&\notag\\
  \left[\alpha^{(J)}_{lm} +\tilde{\alpha}^{(J)}_{lm} \cos\theta + \hat{\alpha}^{(J)}_{lm} \cos^2\theta \right]
  \sin\theta \ {Y}^{lm}(\theta)_{,\theta}- i m \left[\beta^{(J)}_{lm}+\tilde{\beta}^{(J)}_{lm} \cos\theta +\hat{\beta}^{(J)}_{lm}
    \cos^2\theta \right]Y^{lm}(\theta) & \notag\\ 
  +\left[\eta^{(J)}_{lm} + \tilde{\eta}^{(J)}_{lm} \cos\theta\right]\sin^2\theta Y^{lm}(\theta) +
  \left[\xi^{(J)}_{lm}+\tilde{\xi}^{(J)}_{lm} \cos\theta \right] \sin\theta X^{lm}(\theta) +
  \left[\gamma^{(J)}_{lm}+\tilde{\gamma}^{(J)}_{lm} cos\theta \right]\sin^2\theta W^{lm}(\theta)&=0 \,
\label{eq:einst-second-group_A}\\
&\notag\\
-\left[\beta^{(J)}_{lm}+\tilde{\beta}^{(J)}_{lm} cos\theta + \hat{\beta}^{(J)}_{lm} \cos^2\theta +
    \tilde{\Delta}^{(J)}_{lm} \sin^2\theta \right]\sin\theta \ {Y}^{lm}(\theta)_{,\theta}& \notag\\
  - i m
  \left[\alpha^{(J)}_{lm}+\tilde{\alpha}^{(J)}_{lm} \cos\theta +\hat{\alpha}^{(J)}_{lm} \cos^2\theta + \Delta^{(J)}_{lm}
    \sin^2\theta \right]Y^{lm}(\theta) -\left[\zeta^{(J)}_{lm} + \tilde{\zeta}^{(J)}_{lm}
    \cos\theta \right]\sin^2\theta \ Y^{lm}(\theta)   & \notag\\
   -\left[\gamma^{(J)}_{lm}l+\tilde{\gamma}^{(J)}_{lm} \cos\theta \right] \sin\theta X^{lm}(\theta) +
  \left[\xi^{(J)}_{lm}+\tilde{\xi}^{(J)}_{lm} \cos\theta \right]\sin^2 \theta W^{lm}(\theta)&=0 \,,
\label{eq:einst-second-group_B}\\
&\notag\\
\left[f_{lm}+\tilde{f}_{lm} \cos\theta \right]\sin\theta \ {Y}^{lm}(\theta)_{,\theta}  +i m
  \left[g_{lm}+\tilde{g}_{lm} \cos\theta \right]Y^{lm}(\theta)+k_{lm} \sin^2\theta \ Y^{lm}(\theta) & \notag\\
   + \left[s_{lm}+ \hat{s}_{lm} \cos^2\theta \right] \frac{X^{lm}(\theta)}{\sin\theta}+
  \left[t_{lm}+\hat{t}_{lm} \cos^2\theta \right]W^{lm}(\theta)=&0 \,,
\label{eq:einst-third-group_A}\\
&\notag\\
\left[g_{lm}+\tilde{g}_{lm} \cos\theta \right]\sin\theta \ {Y}^{lm}(\theta)_{,\theta}  -i m \left[f_{lm}+
    \tilde{f}_{lm} \cos\theta \right]Y^{lm}(\theta)+\hat{k}_{lm}\sin^2\theta \ Y^{lm}(\theta) &\notag\\
   - \left[t_{lm}+ \hat{t}_{lm} \cos^2\theta\right] \frac{X^{lm}(\theta)}{\sin\theta}+
  \left[s_{lm}+\hat{s}_{lm} \cos^2\theta\right]W^{lm}(\theta)&=0\,, 
\label{eq:einst-third-group_B}
\end{align}
\end{widetext}
where in Eq.~\eqref{eq:einst-first-group}, $I=0,1,2,3$ correspond to the components of Einstein's field equations
behaving as scalars under rotations, and $I=4$ corresponds to the scalar field equation; $J=0,1$ in
Eqs.~\eqref{eq:einst-second-group_A}, \eqref{eq:einst-second-group_B} correspond to the components of Einstein's field
equations behaving as vectors under rotations; and Eqs.~\eqref{eq:einst-third-group_A}, \eqref{eq:einst-third-group_B},
correspond to the components of Einstein's field equations behaving as rank-two tensors under rotations. We have defined
\begin{align}
& X^{lm}(\theta,\varphi)\equiv 2 Y^{lm}_{,\theta \varphi} - 2 \frac{\cos\theta}{\sin\theta}Y^{lm}_{,\varphi} \\
  & W^{lm}(\theta,\varphi)\equiv -2 \frac{\cos\theta}{\sin\theta}Y_{,\theta} - l(l+1) Y^{lm}
  - 2 \frac{Y^{lm}_{,\varphi\varphi}}{\sin^2 \theta}\,.
\label{eq:XW-usual}
\end{align}
The coefficients
\begin{itemize}
\item $A^{(I)}_{lm}$, $\alpha^{(J)}_{lm}$, $\beta^{(J)}_{lm}$, $ s_{lm}$, $ t_{lm}$ contain both zero-th order and
  second order in the spin terms;
\item ${\tilde A}^{(I)}_{lm}$, $C^{(I)}_{lm}$, $B^{(I)}_{lm}$, ${\tilde\alpha}^{(J)}_{lm}$,
${\tilde\beta}^{(J)}_{lm}$, ${\eta}^{(J)}_{lm}$, ${\xi}^{(J)}_{lm}$, ${\gamma}^{(J)}_{lm}$, ${
  \zeta}^{(J)}_{lm}$, $f_{lm}$, $ g_{lm}$ are of order $O({\bar a})$;
\item ${\hat A}^{(J)}_{lm}$, ${\tilde C}^{(J)}_{lm}$, ${E}^{(J)}_{lm}$, ${\tilde B}^{(J)}_{lm}$, ${D}^{(J)}_{lm}$,
  ${\hat \alpha}^{(J)}_{lm}$, ${\hat \beta}^{(J)}_{lm}$, ${\tilde \eta}^{(J)}_{lm}$, ${\tilde \xi}^{(J)}_{lm}$, ${\tilde
  \gamma}^{(J)}_{lm}$,${\tilde \zeta}^{(J)}_{lm}$, ${\Delta}^{(J)}_{lm}$, ${\tilde \Delta}^{(J)}_{lm}$, ${\tilde
  f}^{(J)}_{lm}$, ${\tilde g}^{(J)}_{lm}$, ${k}^{(J)}_{lm}$, ${\hat k}^{(J)}_{lm}$, ${\hat s}^{(J)}_{lm}$, ${\hat
  t}^{(J)}_{lm}$ are of the second order in the spin.
\end{itemize}
All of them are linear combinations of the perturbation functions $h^{lm}_0(r)$, $h^{lm}_1(r)$, $H^{lm}_0(r)$,
$H^{lm}_1(r)$, $H^{lm}_2(r)$, $K^{lm}(r)$, $\Phi^{lm}(r)$ and their derivatives, with coefficients that depend on $l$
but not on $m$. Their explicit expansions in the coupling parameter $\zeta$, up to $O(\zeta^6)$, are given in the
Supplemental Material~\cite{notebook}.
 
We project Eqs.~\eqref{eq:einst-first-group}-\eqref{eq:einst-third-group_B} on the complete set of tensor spherical
harmonics, as in~\cite{Kojima1992}, decoupling the angular variables:
\begin{widetext}
\begin{flalign}
  & A^{(I)}_{lm}+\hat{A}^{(I)}_{lm}\left[Q_{lm}^2+Q_{l+1\,m}^2\right]+ \tilde{B}^{(I)}_{lm}
  \left[l Q^2_{l+1\,m}-(l+1)Q^2_{lm}\right] + i m C^{(I)}_{lm} + m^2 E^{(I)}_{lm} +Q_{lm}
  \Big\{ \left[\tilde{A}^{(I)}_{l-1\,m}+(l-1)B^{(I)}_{l-1\,m}\right] \notag\\
  & + i m \left[\tilde{C}^{(I)}_{l-1\,m}+(l-1)D^{(I)}_{l-1\,m}\right]\Big\} +Q_{l+1\,m}\Big\{\left[\tilde{A}^{(I)}_{l+1\,m}
    -(l+2)B^{(I)}_{l+1\,m}\right]+i m \left[\tilde{C}^{(I)}_{l+1\,m}-(l+2)D^{(I)}_{l+1\,m}\right]\Big\}\notag\\
  &+ Q_{lm} Q_{l-1\,m}\left[\hat{A}^{(I)}_{l-2\,m}+(l-2)\tilde{B}^{(I)}_{l-2\,m}\right] + Q_{l+1\,m}Q_{l+2\,m}
  \left[\hat{A}_{l+2\,m}^{(I)}-(l+3)\tilde{B}^{(I)}_{l+2\,m}\right] = 0 \,,&&
\label{eq:eq1-2nd}
\end{flalign}
\begin{flalign}
  & l(l+1)\alpha^{(J)}_{lm}- i m \left[\tilde{\beta}^{(J)}_{lm} + \zeta^{(J)}_{lm}-(l-1)(l+2)\xi^{(J)}_{lm}\right]+
  \left[(l+1)(l-2)Q_{lm}^2+l(l+3)Q^2_{l+1\,m}\right]\hat{\alpha}^{(J)}_{lm} \notag\\
  & + m^2 \Delta^{(J)}_{lm} +\left[l Q^2_{l+1\,m}-(l+1)Q_{lm}^2\right]\tilde{\eta}^{(J)}_{lm}+ \left[2 m^2
    +Q^2_{lm}(l+1)(l^2-l+4)-Q^2_{l+1\,m} l (l^2+3 l+6)\right]\tilde{\gamma}^{(J)}_{lm} \notag\\
  & + Q_{lm} \Big\{(l-1)(l+1)\tilde{\alpha}^{(J)}_{l-1\,m} -(l+1) \eta^{(J)}_{l-1\,m}+(l-2)(l-1)(l+1)
  \gamma^{(J)}_{l-1\,m} && \notag\\
  & \ \ \ \ - i m \left[2 \hat{\beta}^{(J)}_{l-1\,m}+(l-1)\tilde{\Delta}^{(J)}_{l-1\,m}+ \tilde{\zeta}^{(J)}_{l-1\,m}-
    (l-2)(l+3)\tilde{\xi}^{(J)}_{l-1\,m}\right] \Big\}\notag\\
& + Q_{l+1\,m} \Big\{l(l+2)\tilde{\alpha}^{(J)}_{l+1\,m} +l \eta^{(J)}_{l+1\,m}-l (l+2)(l+3)\gamma^{(J)}_{l+1\,m} && \notag \\
  & \ \ \ \ - i m \left[2 \hat{\beta}^{(J)}_{l+1\,m}-(l+2)\tilde{\Delta}^{(J)}_{l+1\,m}+ \tilde{\zeta}^{(J)}_{l+1\,m}
    -(l-2)(l+3)\tilde{\xi}^{(J)}_{l+1\,m}\right] \Big\} \notag\\
  & + Q_{l-1\,m}Q_{lm}\Big\{(l-2)(l+1)\hat{\alpha}^{(J)}_{l-2\,m}-(l+1)\tilde{\eta}^{(J)}_{l-2\,m} +(l-2)(l+1)(l-3)
  \tilde{\gamma}^{(J)}_{l-2\,m}\Big\} \notag\\
  & + Q_{l+1\,m}Q_{l+2\,m}\Big\{l(l+3)\hat{\alpha}^{(J)}_{l+2}+l \tilde{\eta}_{l+2\,m}^{(J)}-l(l+3)(l+4)\
  \tilde{\gamma}^{(J)}_{l+2\,m)} \Big\} = 0\,,&&
\label{eq:eq2-2nd}
\end{flalign}
\begin{flalign}
  & l(l+1)\beta^{(J)}_{lm} + i m \left[\tilde{\alpha}^{(J)}_{lm} + \eta^{(J)}_{lm} + (l-1)(l+2)\gamma^{(J)}_{lm}\right]+
  \left[(l+1)(l-2)Q_{lm}^2+l(l+3)Q^2_{l+1\,m}\right]\hat{\beta}^{(J)}_{lm}  +\big[l^2 Q^2_{l+1\,m}\notag\\
    & + (l+1)^2 Q_{lm}^2\big]\tilde{\Delta}^{(J)}_{lm} +\left[l Q^2_{l+1\,m}-(l+1) Q_{lm}^2\right]\tilde{\zeta}^{(J)}_{lm}-
  \left[2 m^2 +Q^2_{lm}(l+1)(l^2-l+4)-Q^2_{l+1\,m} l (l^2+3l+6)\right]\tilde{\xi}^{(J)}_{lm} \notag\\
& + Q_{lm} \Big\{(l-1)(l+1)\tilde{\beta}^{(J)}_{l-1\,m} -(l+1) \zeta^{(J)}_{l-1\,m}-(l-2)(l-1)(l+1)\xi^{(J)}_{l-1\,m} \notag\\
  & \ \ \ \ + i m \left[2 \hat{\alpha}^{(J)}_{l-1\,m}-(l+1)\Delta^{(J)}_{l-1\,m}+ \tilde{\eta}^{(J)}_{l-1\,m}
    +(l-2)(l+3)\tilde{\gamma}^{(J)}_{l-1\,m}\right] \Big\}\notag\\
  & + Q_{l+1\,m} \Big\{l(l+2)\tilde{\beta}^{(J)}_{l+1\,m} +l \zeta^{(J)}_{l+1\,m}+l (l+2)(l+3)\xi^{(J)}_{l+1\,m}
  + i m \left[2 \hat{\alpha}^{(J)}_{l+1\,m}+l\Delta^{(J)}_{l+1\,m}+ \tilde{\eta}^{(J)}_{l+1\,m}+
    (l-2)(l+3)\tilde{\gamma}^{(J)}_{l+1\,m}\right] \Big\} \notag\\
  & + Q_{l-1\,m}Q_{lm}\Big\{(l-2)(l+1)\hat{\beta}^{(J)}_{l-2\,m}-(l+1)\tilde{\zeta}^{(J)}_{l-2\,m} -
  (l-2)(l+1)(l-3) \tilde{\xi}^{(J)}_{l-2\,m}-(l-2)(l+1)\tilde{\Delta}^{(J)}_{l-2\,m}\Big\} \notag\\
  & + Q_{l+1\,m}Q_{l+2\,m}\Big\{l(l+3)\hat{\beta}^{(J)}_{l+2\,m}+l \tilde{\zeta}_{l+2\,m}^{(J)}+l(l+3)(l+4)\tilde{\xi}^{(J)}_{l+2\,m}
  -l(l+3)\tilde{\Delta}^{(J)}_{l+2\,m} \Big\} = 0\,,&&
\label{eq:eq3-2nd}
\end{flalign}
\begin{flalign}
  & l(l-1)(l+1)(l+2) s_{lm} - i m (l-1)(l+2)f_{lm} + \left[2 m^2 + Q_{lm}^2 (l+1)(l^2-l+4)-Q^2_{l+1\,m}l(l^2+3l+6)\right]
  \tilde{g}_{lm} \notag\\
& + \left[2 m^2 -l(l+1) +(l+1)(l+2) Q_{lm}^2 + l(l-1) Q^2_{l+1\,m}\right]\hat{k}_{lm} \notag\\
  & + \Big\{8 m^2-2 l(l+1)-Q_{lm}^2 (l+1)\left[4(l-2)-l(l+1)(l+4)\right]-Q_{l+1\,m}^2 l \left[4(l+3)-l(l+1)(l-3)\right]
  \Big\}\hat{s}_{lm} \notag\\
  & - Q_{lm}\Big\{(l-1)(l+1)(l+2) g_{l-1\,m}+ i m \left[(l-3)(l+2)\tilde{f}_{l-1\,m}- 2(l+2) k_{l-1\,m} + 4 (l-2)(l+2)
    \hat{t}_{l-1\,m}\right] \Big\} \notag\\
  & + Q_{l+1\,m}\Big\{l(l-1)(l+2)g_{l+1\,m} - i m \left[(l-1)(l+4) \tilde{f}_{l+1\,m}+2 (l-1) k_{l+1\,m} +4 (l-1)(l+3)
    \hat{t}_{l+1\,m} \right] \Big\} \notag\\
& + Q_{l-1\,m}Q_{lm} (l+1)(l+2)\Big\{ -(l-2)\tilde{g}_{l-2\,m} + \hat{k}_{l-2\,m}+(l-3)(l-2) \hat{s}_{l-2\,m}\Big\} \notag\\
& + Q_{l+1\,m}Q_{l+2\,m} l (l-1)\Big\{ (l+3)\tilde{g}_{l+2\,m} + \hat{k}_{l+2\,m}+(l+3)(l+4) \hat{s}_{l+2\,m}\Big\} =0\,, &&
\label{eq:eq4-2nd}
\end{flalign}
\begin{flalign}
  & 0 = l(l-1)(l+1)(l+2) t_ {l}  + i m (l-1)(l+2)g_{lm} + \left[2 m^2 + Q_{lm}^2 (l+1)(l^2-l+4)-Q^2_{l+1\,m}l(l^2+3l+6)
    \right]\tilde{f}_{lm} \notag\\
& + \left[2 m^2 -l(l+1) +(l+1)(l+2) Q_{lm}^2 + l(l-1) Q^2_{l+1\,m}\right]k_{lm} \notag\\
  & + \Big\{8 m^2-2 l(l+1)-Q_{lm}^2 (l+1)\left[4(l-2)-l(l+1)(l+4)\right]-Q_{l+1\,m}^2 l \left[4(l+3)-l(l+1)(l-3)\right]
  \Big\}\hat{t}_{lm} \notag\\
  & - Q_{lm}\Big\{(l-1)(l+1)(l+2) f_{l-1\,m}- i m \left[(l-3)(l+2)\tilde{g}_{l-1\,m}- 2(l+2) \hat{k}_{l-1\,m} +
    4 (l-2)(l+2) \hat{s}_{l-1\,m}\right] \Big\} \notag\\
  & + Q_{l+1\,m}\Big\{l(l-1)(l+2)f_{l+1} + i m \left[(l-1)(l+4) \tilde{g}_{l+1\,m}+2 (l-1) \hat{k}_{l+1\,m}
    +4 (l-1)(l+3)\hat{s}_{l+1\,m} \right] \Big\} \notag\\
& + Q_{l-1\,m}Q_{lm} (l+1)(l+2)\Big\{ -(l-2)\tilde{f}_{l-2\,m} + k_{l-2\,m}+(l-3)(l-2) \hat{t}_{l-2\,m}\Big\} \notag\\
  & + Q_{l+1\,m}Q_{l+2\,m} l (l-1)\Big\{ (l+3)\tilde{f}_{l+2\,m} + k_{l+2\,m}+(l+3)(l+4) \hat{t}_{l+2\,m}\Big\}\,,&&
\label{eq:eq5-2nd}
\end{flalign}
\end{widetext}
where $l\ge2$ and
\begin{equation}
Q_{lm}=\sqrt{\frac{(l-m)(l+m)}{(2l-1)(2l+1)}}\,.
\label{eq:defQ}
\end{equation}
As discussed in Sec.~\ref{sec:secondorder}, for polar-led perturbations with $l=2$ some of the terms in 
Eqs.~\eqref{eq:eq1-2nd} - \eqref{eq:eq5-2nd} vanish, thus they reduce to Eqs. \eqref{eq:eq1PL}-\eqref{eq:eq5bPL}.
%


\bibliographystyle{utphys}
\bibliography{bibliography}

\end{document}